\documentclass[12pt,preprint]{aastex}

\usepackage{emulateapj5}
\usepackage{aastexug}
\usepackage{amsmath}
\usepackage{graphicx}
\usepackage{subfigure}
\usepackage{apjfonts}

\begin{document}
\def\gapprox{\mathrel{\vcenter{\offinterlineskip \hbox{$>$}
    \kern 0.3ex \hbox{$\sim$}}}}
\def\lapprox{\mathrel{\vcenter{\offinterlineskip \hbox{$<$}
    \kern 0.3ex \hbox{$\sim$}}}}

\newcommand{\Dt}[0]{\bigtriangleup t}
\newcommand{\Dx}[0]{\bigtriangleup x}
\newcommand{\E}{\mathcal{E}}
\newcommand{\D}{\bigtriangleup}
\newcommand{\beq}{\begin{equation}}
\newcommand{\eeq}{\end{equation}}
\newcommand{\mm}[2]{\textrm{minmod}\left({#1},{#2}\right)}
\newcommand{\sign}{\textrm{sign}}
\newcommand{\nf}{\mathcal{F}}
\newcommand{\pfrac}[2]{\left(\frac{#1}{#2}\right)}

\title{Implementation of the Shearing Box Approximation in Athena}

\author{James M. Stone}
\affil{Department of Astrophysical Sciences, Princeton University, Princeton,
NJ 08544}
\author{and Thomas A. Gardiner}
\affil{Sandia National Laboratories, Albuquerque, NM 87185-1189}

\begin{abstract}
We describe the implementation of the shearing box approximation
for the study of the dynamics of accretion disks in the Athena
magnetohydrodynamics (MHD) code.  Second-order Crank-Nicholson time
differencing is used for the Coriolis and tidal gravity source terms that
appear in the momentum equation for accuracy and stability.  We show this
approach conserves energy for epicyclic oscillations in hydrodynamic flows
to round-off error.
In the energy equation, the tidal gravity source terms are differenced
as the gradient of an effective potential in a way which guarantees that
total energy (including the gravitational potential energy) is also conserved
to round-off error.  We introduce an orbital advection algorithm for
MHD based on constrained transport to preserve the divergence-free
constraint on the magnetic field.  This algorithm removes the orbital
velocity from the time step constraint, and makes the truncation error
more uniform in radial position.  Modifications to the
shearing box boundary conditions applied at the radial boundaries are
necessary to conserve the total vertical magnetic flux.  In principle
similar corrections are also required to conserve mass, momentum and energy,
however in practice we find the orbital advection method conserves these
quantities to better than 0.03\% over hundreds of orbits.  The algorithms
have been applied to studies of the nonlinear regime of the MRI in very
wide (up to 32 scale heights) horizontal domains.

\end{abstract}


\section{Introduction}

Numerical methods are an important tool for studying the nonlinear
gas dynamics in accretion flows.  For example, much of what we have
learned about the saturation of the magnetorotational instability
(MRI) has come from magnetohydrodynamic (MHD) simulations (Balbus \&
Hawley 2003).  A large fraction of such calculations adopt a local
approximation (Hill 1878) termed the ``shearing box", first introduced
in studies of the MRI by Hawley et al. (1995, hereafter HGB).  In this
approximation, the equations of motion are written in a local, Cartesian
reference frame co-rotating with the disk at some arbitrary radius $r_0$.
The approximation is valid provided the linear extent of the domain under
study is small compared to $r_0$.  The shearing box approximation has
limitations (Regev \& Umurhan 2008), in particular it cannot be used
to calculate important global properties of the disk like the net mass
accretion rate, radial density and temperature profiles, or the spectrum
of emitted radiation.  Nevertheless it has provided a useful laboratory
for the study of important questions related to the local dynamics of
accretion disks (Balbus 2003).  For example, recent studies that use
the shearing box include the effect of microscopic diffusivities on
the saturation level of MHD turbulence driven by the MRI (Fromang et
al. 2007; Lesur \& Longaretti 2007; Simon \& Hawley 2009), the saturation
of the MRI in radiation pressure dominated disks (Hirose et al. 2008),
and the effects of non-ideal MHD and dust on the properties of the MRI
in protostellar disks (Turner \& Sano 2008; Ilgner \& Nelson 2008).

The shearing box approximation requires source terms be added to the
momentum and energy equations, and special boundary conditions be used
in the radial direction.  Most of the simulations of the MRI presented
to date have used operator-split methods like ZEUS (e.g. HGB; Stone et
al. 1996), or simple finite-difference methods like the PENCIL code
(e.g. Johansen et al. 2009).  In both cases, adding the shearing box
source terms is straightforward (HGB).  More recently, higher-order
Godunov methods that adopt the conservative form have begun to be applied
to studies of accretion flows in the shearing box (Shen et al. 2006;
Fromang \& Papaloizou 2007; Bodo et al. 2008; Piontek et al. 2009; Simon
et al. 2009; Tilley et al. 2009; Gressel 2010).

In this paper, we provide a detailed description of the algorithmic
extensions for the shearing box approximation in Athena, a recently
developed higher-order Godunov code for astrophysical MHD.  The basic
MHD algorithms in Athena are documented in Gardiner \& Stone (2005a;
2008), and details of the implementation and tests of the methods are
given in Stone et al. (2008, hereafter SGTHS) and Stone \& Gardiner
(2009).  Currently there are two versions of Athena, one implemented
in C and the other in Fortran.  An extension of the Fortran version
for the shearing box has already been used for new studies of the MRI
(Simon \& Hawley 2009; Simon et al. 2009).  The  numerical algorithms
in the C version are slightly different.  They were first introduced
by Gardiner \& Stone (2005b), and used to study hydrodynamic shearing
waves by Shen et al. (2006) and Balbus \& Hawley (2006).  They have
the advantage of preserving the energy integral in epicyclic motion to
round-off, conserving the total (including gravitational potential)
energy to round-off, and producing virtually no aliasing of trailing
into leading waves.  We emphasize, however, that since epicyclic motion
is destabilized by a weak magnetic field (Balbus \& Hawley 1991), the
accuracy of this approach can only be quantitatively demonstrated for
hydrodynamic flows.  Although similar methods were adopted and extended
by  Gressel \& Ziegler (2007), this paper provides the first comprehensive
description of the algorithms in the C version of Athena.  The C version
is also being used for new studies of the MRI (Davis et al. 2010).

In addition to methods for the shearing box source terms, we also
describe the implementation of an orbital advection algorithm for MHD
in Athena.  Orbital advection can greatly increase the efficiency of
calculations in domains that span more than one scale height in the
radial direction (so that the difference in the orbital speed across
the domain is supersonic), since the time step constraint for stability
does not depend on the magnitude of the local orbital velocity, but only
on the amplitude of the fluctuations in the velocity around this value.
Moreover, orbital advection can improve the accuracy of the integration by
making the truncation error more uniform in radius (Johnson et al. 2008,
hereafter JGG).  Orbital advection was first introduced for hydrodynamic
studies of disks by Masset (2000) in the FARGO code.  More recently,
JGG have described an extension of orbital advection methods to MHD for
ZEUS-type codes.  Johansen et al. (2009) have described the extension of
the PENCIL code with orbital advection using Fourier transform methods.
The method we have implemented in Athena is quite different from
these previous approaches.  In particular, we update the magnetic
field using constrained transport (Evans \& Hawley 1988) to guarantee
the divergence-free constraint is enforced to machine precision, using
an effective electric field produced by the orbital motion.  This greatly
simplifies the method.

The organization of this paper is as follows.  In the following section,
we catalog the basic equations solved by Athena in the shearing box
approximation.  In \S 3 we describe our implementation of the source terms
in these equations, including tests of our methods.  In \S 4
we describe the shearing box boundary conditions for the conservative
variables, including issues associated with parallelization of the
algorithms with MPI.  Finally, in \S 5 we describe our orbital advection
method, along with tests.  We present preliminary
simulations of the MRI in very wide horizontal domains and summarize in
\S 6.

\section{Basic Equations}

The local shearing box approximation adopts a frame of reference
located at radius $r_{0}$ corotating with the disk at orbital frequency
$\Omega_{0}=\Omega(r_{0})$.  In this frame, the equations of MHD are
written in a Cartesian coordinate system $(x,y,z)$ that has unit vectors
${\bf \hat{i}}$, ${\bf \hat{j}}$, and ${\bf \hat{k}}$ as
\begin{eqnarray}
\frac{\partial \rho}{\partial t} +
{\bf\nabla\cdot} [\rho{\bf v}] & = & 0,
\label{eq:cons_mass} \\
\frac{\partial \rho {\bf v}}{\partial t} +
{\bf\nabla\cdot} \left[\rho{\bf vv} + {\sf T}\right] & = &
\rho \Omega_{0}^{2}(2qx{\bf \hat{i}} - z{\bf \hat{k}}) - 2\Omega_{0} {\bf \hat{k}} \times \rho {\bf v},
\label{eq:cons_momentum} \\
\frac{\partial E}{\partial t} +
\nabla\cdot \left[E{\bf v} + {\sf T} \cdot {\bf v}\right] & = &
\Omega_{0}^{2}\rho{\bf v} \cdot (2qx{\bf \hat{i}} - z{\bf \hat{k}}) ,
\label{eq:cons_energy} \\
\frac{\partial {\bf B}}{\partial t} -
{\bf\nabla} \times \left({\bf v} \times {\bf B}\right) & = & 0,
\label{eq:induction}
\end{eqnarray}
where ${\sf T}$ is the total stress tensor
\begin{equation}
  {\sf T} = (p + B^{2}/2){\sf I} - {\bf B}{\bf B},
\label{eq:stresstensor}
\end{equation}
$p$ is the gas pressure, $E$ is the sum of the internal, kinetic, and
magnetic energy densities
\begin{equation}
  E = \frac{p}{\gamma -1} + \frac{1}{2}\rho v^{2} + \frac{B^{2}}{2},
\label{eq:totalenergy}
\end{equation}
and $v^{2} = {\bf v} \cdot {\bf v}$, 
$B^{2} = {\bf B} \cdot {\bf B}$.  The shear parameter $q$ is defined as
\begin{equation}
 q = - \frac{1}{2} \frac{d {\rm ln} \Omega^2}{d {\rm ln} r}
\label{eq:shearparam}
\end{equation}
so that for Keplerian flow $q=3/2$.
The other symbols have their usual meaning.  These equations are written
in units such that the magnetic permeability $\mu=1$, and an equation of
state appropriate to an ideal gas has been assumed in writing equation
\ref{eq:totalenergy}, that is $p=(\gamma -1)e$ (where $\gamma$ is
the ratio of specific heats, and $e$ is the internal energy density).
The techniques described in this paper are easily generalized to other
equations of state.

The first two source terms on the RHS of equation \ref{eq:cons_momentum}
represent the radial and vertical components of the gravitational force in
the rotating frame, while the third term represents the Coriolis force.
In the energy equation \ref{eq:cons_energy}, the two source terms on
the RHS represent the work done by the radial and vertical components
of gravity in the rotating frame.  Most of the challenges associated
with the addition of the shearing box source terms to Godunov schemes
like Athena are related to the tidal gravity and Coriolis terms which
act in the orbital ($x-y$) plane, therefore in the discussion below
we describe the algorithms for the source terms in the horizontal and
vertical directions separately.

The equations of motion in the shearing box admit a simple
equilibrium solution representing uniform orbital motion,
\begin{equation}
{\bf v}_{K} = -q\Omega_{0} x {\bf \hat{j}}.
\label{eq:shearflow}
\end{equation}
Note this velocity is time-independent, varies only in the $x-$direction,
and only the $y-$component is non-zero.  These properties are important
for the orbital advection scheme described in \S5.

\section{Shearing Box Source Terms}

One approach to implementing source terms in Godunov schemes like Athena
is to operator split them from the flux divergence terms, resulting
in a system of ODEs that can be integrated with any number of methods.
For example, Simon et al. (2009) have successfully used this approach
to study the saturation of the MRI with the Fortran version of Athena.
However, in studies of decaying {\em hydrodynamic} turbulence in
the shearing box (Shen et al. 2006), we found the kinetic energy in
velocity fluctuations could {\em increase} at very late times, rather
than decay monotonically to zero.  We traced the source of this growth
to an artificial amplification of epicyclic motions by the truncation
error associated with the source terms in the momentum equations.  As we
show below, in hydrodynamics it is possible to develop a discretization
that will instead conserve this energy to round-off error.  While this
approach may not provide significant improvement for MHD flows, since
epicyclic motion is unstable with weak magnetic fields, we show below
it is more accurate for hydrodynamics, and therefore we have adopted
it as the default algorithm in both hydrodynamics and MHD.  The methods
for integrating the source terms described below were first introduced
by Gardiner \& Stone (2005b), subsequently Gressel \& Ziegler (2007)
have adopted and expanded upon this approach.

To describe our methods, it is useful to define the $x-$ and $y-$components
of the momentum density fluctuations
\beq
m = \rho v_{x}
\label{eq:x-mom-fluct}
\eeq
\beq
n = \rho (v_{y} + q\Omega_{0}x) 
\label{eq:x-mom-fluct}
\eeq
To investigate the properties of epicyclic motion in a 
stress-free medium, we consider the case in which the velocity
fluctuation is a function of time and
rewrite the
momentum equation \ref{eq:cons_momentum} using these variables, giving
\begin{equation}
\frac{\partial m}{\partial t} = 2 \Omega_{0} n
\label{eq:oned-m}
\end{equation}
\begin{equation}
\frac{\partial n}{\partial t} = (q-2)\Omega_{0} m
\label{eq:oned-n}
\end{equation}
Multiplying equation \ref{eq:oned-m} by $m$, equation
\ref{eq:oned-n} by $-2n/(q-2)$, and adding gives
\begin{equation}
 \frac{\partial}{\partial t} \left( m^2 + \frac{2}{2-q}n^2 \right) = 0
\end{equation}
Thus, there is a conserved energy which for Keplerian flow $(q=3/2)$
is $E_{\rm epi} = [m^2 + 4n^2]/(2\rho)$.

We would like the numerical discretization of the momentum equation to
conserve $E_{\rm epi}$ exactly.  Remarkably,
nothing more complicated than Crank-Nicholson time differencing is
required.  Following the usual convention, we discretize time into $N$
non-uniform steps, and use a superscript to denote the time level
of any quantity, with the timestep $\delta t$ defined as $\delta t = t^{n+1}
-t^{n}$.  We define time difference and averaging operators as
\begin{equation}
[x] = x^{n+1} - x^{n} 
\label{eq:t-difference}
\end{equation}
\begin{equation}
\bar{x} = \frac{1}{2} \left( x^{n+1} + x^{n} \right)
\label{eq:t-average}
\end{equation}
respectively.  Then, the Crank-Nicholson time difference formulae for equations
\ref{eq:oned-m} and \ref{eq:oned-n} are
\begin{equation}
[m] = 2 \Omega_{0} \delta t \bar{n}
\end{equation} 
\begin{equation}
[n] = (q-2)\Omega_{0} \delta t \bar{m}
\end{equation}
By multiplying the first of these equations by $\bar{m}$, the second by
$-2\bar{n}/(q-2)$, and adding gives, after some manipulation 
\begin{equation}
[m^{2} + \frac{2}{2-q}n^{2}] = 0.
\end{equation} 
Thus, Crank-Nicholson discretization conserves the energy integral for
epicyclic oscillations in a discrete sense.  There is a simple physical
interpretation of this result.  The Coriolis force ${\bf F}_c = (- 2
{\bf \Omega} \times {\bf v})$, hence ${\bf v}\cdot {\bf F}_c =0$ and the
Coriolis force can do no work.  Without care, the discretized Coriolis
force may not be orthogonal to the velocity (in a time averaged sense).
This can lead to unphysical growth or decay of the energy in epicyclic
motion.  Our tests have revealed that a forward Euler discretization
leads to a growing amplitude for epicyclic oscillations, while a backward
Euler discretization leads to a decaying amplitude.  Fortunately, the
average of the two (Crank-Nicholson differencing), conserves
the energy.  Similarities
can be drawn between integrating fluid motion due to the
Coriolis force, and
particle orbits in a central potential, e.g. forward Euler applied to the
latter leads to outward spiraling rather than closed orbits.

\subsection{Source Terms in the Momentum Equation}

We now develop a finite-volume discretization of the momentum equation
\ref{eq:cons_momentum}, using a Crank-Nicholson time discretization of
the tidal gravity and Coriolis source terms.  We consider only
the $x-$ and $y-$components of this equation.  Written in terms of the
momentum density fluctuations, $m$ and $n$ respectively, these equations are
\begin{equation}
\frac{\partial m}{\partial t} + \nabla \cdot {\bf F}_{m} = 2 \Omega_{0} n
\label{eq:multid-m}
\end{equation} 
\begin{equation}
\frac{\partial n}{\partial t} + \nabla \cdot {\bf F}_n = (q-2)\Omega_{0} m
\label{eq:multid-n}
\end{equation} 
where ${\bf F}_m$ and ${\bf F}_n$ are vectors whose components are
the fluxes of the momentum density fluctuations in each direction.
These are related to the vector of fluxes of the corresponding 
components of the momentum density, ${\bf F}_{\rho v_x}$ and
${\bf F}_{\rho v_y}$ respectively, via
\beq
{\bf F}_{m} = {\bf F}_{\rho v_x}
\label{eq:m-flux}
\eeq
\beq
{\bf F}_{n} = {\bf F}_{\rho v_y} + q\Omega_{0} x {\bf F}_{\rho}
\label{eq:n-flux}
\eeq
We now adopt a finite-volume discretization of equations \ref{eq:multid-m}
and \ref{eq:multid-n} (see \S3 in SGTHS for a discussion of this
approach).  We use indices $(i,j,k)$ to denote spatial locations on
a discrete grid with cell centered locations $(x_i, y_j, z_k)$.
Half integer indices are used to denote the appropriate cell
interfaces.  The dependent variables stored on this mesh $m_{i,j,k}^n$
and $n_{i,j,k}^n$ are understood to be volume averaged values in the
sense of equation 12 in SGTHS.  Integrating over a timestep $\delta t =
t^{n+1}-t^{n}$ and over the volume of a cell, equations \ref{eq:multid-m}
and \ref{eq:multid-n} become
\beq
\frac{(m_{i,j,k}^{n+1} - m_{i,j,k}^n)}{\delta t} +
\overline{\left( {\bf \nabla} \cdot {\bf F}_{m} \right)}_{i,j,k} =
2 \Omega_{0} \frac{(n_{i,j,k}^n + n_{i,j,k}^{n+1})}{2}
\label{eq:m_cons_evol}
\eeq
\beq
\frac{(n_{i,j,k}^{n+1} - n_{i,j,k}^n)}{\delta t} +
\overline{\left( {\bf \nabla} \cdot {\bf F}_{n} \right)}_{i,j,k} =
(q-2)\Omega_{0} \frac{(m_{i,j,k}^n + m_{i,j,k}^{n+1})}{2}
\label{eq:n_cons_evol}
\eeq
In a finite volume approach, the time- and volume-average of the flux
divergence (the second term in both equations) would be rewritten, using
the divergence theorem, as the difference of the time- and area-averaged
fluxes at each of the cell faces, e.g. equations 11-15 in SGTHS.
In a Godunov scheme, these fluxes are computed with a Riemann solver.
For brevity, we have suppressed expanding the flux divergence terms into
these differences, with the understanding that our notation is meant
to represent these terms.

Solving equation \ref{eq:n_cons_evol} for $(n_{i,j,k}^{n} + n_{i,j,k}^{n+1})$
and substituting into equation \ref{eq:m_cons_evol} gives,
after some manipulation
\begin{eqnarray}
m_{i,j,k}^{n+1} &=& m_{i,j,k}^{n} - 
\delta t \overline{\left( {\bf \nabla} \cdot {\bf F}_{m} \right)}_{i,j,k}
\nonumber \\
&+& \left(\frac{4\Omega_{0} \delta t}{2 + (2-q) (\Omega_{0} \delta t)^2} \right)
\left( \left\{ n_{i,j,k}^n - \frac{\delta t}{2} 
\overline{\left( {\bf \nabla} \cdot {\bf F}_{n} \right)}_{i,j,k}
\right\} \right. \nonumber \\
&+& \left. \frac{(q-2)}{2}\Omega_{0} \delta t 
\left\{ m_{i,j,k}^{n} - \frac{\delta t}{2} 
\overline{\left( {\bf \nabla} \cdot {\bf F}_{m} \right)}_{i,j,k} \right\}
\right)
\label{eq:m_cons_evol_2}
\end{eqnarray}
Similarly, solving equation \ref{eq:m_cons_evol} for
$(m_{i,j,k}^{n} + m_{i,j,k}^{n+1})$ and substituting
into equation \ref{eq:n_cons_evol} gives
\begin{eqnarray}
n_{i,j,k}^{n+1} &=& n_{i,j,k}^n 
- \delta t \overline{\left( {\bf \nabla} \cdot {\bf F}_{n} \right)}_{i,j,k}
\nonumber \\
&+& \left(\frac{2(q-2)\Omega_{0} \delta t} {2 + (2-q) (\Omega_{0} \delta t)^2}\right) 
\left( \left\{ m_{i,j,k}^n - \frac{\delta t}{2} 
\overline{\left( {\bf \nabla} \cdot {\bf F}_{m} \right)}_{i,j,k}
\right\} \right. \nonumber \\
&+& \left. \Omega_{0} \delta t 
\left\{ n_{i,j,k}^n - \frac{\delta t}{2} 
\overline{\left( {\bf \nabla} \cdot {\bf F}_{n} \right)}_{i,j,k} \right\}
\right)
\label{eq:n_cons_evol_2}
\end{eqnarray}

In order to convert these update equations for
$m_{i,j,k}$ and $n_{i,j,k}$ respectively, into update
equations for $(\rho v_{x})_{i,j,k}$ and $(\rho v_{y})_{i,j,k}$ (the discrete
form of the conserved
quantities actually updated in the code), relationships between the discrete
fluxes of these quantities are required.  For any cell $(i,j,k)$ with
zone-center $x$-position $x_{i}$, the relationships between the components of
the volume averaged momentum density fluctuations and the momentum density are
\beq
m_{ijk} = (\rho v_x)_{ijk}
\label{eq:m-discrete-mom}
\eeq
\beq
n_{ijk} = (\rho v_y)_{ijk} + q\Omega_{0} \rho_{ijk} x_{i}
\label{eq:n-discrete-mom}
\eeq
From equations \ref{eq:m-flux} and \ref{eq:n-flux}, the divergence of the
fluxes of these quantities are related via 
\beq
{\bf \nabla} \cdot {\bf F}_{m} = {\bf \nabla} \cdot {\bf F}_{\rho v_x}
\eeq
\beq
{\bf \nabla} \cdot {\bf F}_{n} = {\bf \nabla} \cdot {\bf F}_{\rho v_y}
+ q\Omega_{0} {\bf \nabla} \cdot ( x  {\bf F}_{\rho} )
\eeq
which can be written in a finite volume discretization as
\beq
\overline{\left( {\bf \nabla} \cdot {\bf F}_{m} \right)}_{i,j,k} = 
\overline{\left( {\bf \nabla} \cdot {\bf F}_{\rho v_x} \right)}_{i,j,k}
\label{eq:disc_x_mom_flux_div_rel}
\eeq
\begin{eqnarray}
\overline{\left( {\bf \nabla} \cdot {\bf F}_{n} \right)}_{i,j,k} &=& 
\overline{\left( {\bf \nabla} \cdot {\bf F}_{\rho v_y} \right)}_{i,j,k}
+ q\Omega_{0} x_{i} 
\overline{\left( {\bf \nabla} \cdot {\bf F}_{\rho} \right)}_{i,j,k} 
\nonumber \\
&+& (q/2)\Omega_{0} x_{i} \left( (F_\rho)_{x,i-1/2} + (F_\rho)_{x,i+1/2} \right)
\label{eq:disc_y_mom_flux_div_rel}
\end{eqnarray}
Finally, inserting equations \ref{eq:n-discrete-mom} and
\ref{eq:disc_y_mom_flux_div_rel} into \ref{eq:n_cons_evol_2} gives
\begin{eqnarray}
(\rho v_y)_{i,j,k}^{n+1} &=& (\rho v_y)_{i,j,k}^n - \delta t 
\overline{\left( {\bf \nabla} \cdot {\bf F}_{(\rho v_y)} \right)}_{i,j,k}
\nonumber \\
&-&\frac{q}{2} \Omega_{0} \delta t x_{i} 
\left( (F_\rho)_{x,i-1/2} + (F_\rho)_{x,i+1/2} \right)
\nonumber \\
&+& \left(\frac{2(q-2)\Omega_{0} \delta t} {2 + (2-q) (\Omega_{0} \delta t)^2}\right) 
\left( \left\{ m_{i,j,k}^n - \frac{\delta t}{2} 
\overline{\left( {\bf \nabla} \cdot {\bf F}_{m} \right)}_{i,j,k}
\right\} \right. \nonumber \\
&+& \left. \Omega_{0} \delta t 
\left\{ n_{i,j,k}^n - \frac{\delta t}{2} 
\overline{\left( {\bf \nabla} \cdot {\bf F}_{n} \right)}_{i,j,k} \right\}
\right)
\label{eq:rhovy_cons_evol_3}
\end{eqnarray}
Since the $m_{i,j,k}$ and 
$(\rho v_{x})_{i,j,k}$ 
are identical, the update
relation for the latter is given by equation \ref{eq:m_cons_evol_2}.

Equations \ref{eq:m_cons_evol_2} and \ref{eq:rhovy_cons_evol_3}
above represent the desired finite volume discretization of the
momentum equation, where the divergence of the fluxes of the momentum
fluctuations that appear in these equations are given in terms of the
divergence of the fluxes of the conserved quantities actually returned
by the Riemann solver by equations \ref{eq:disc_x_mom_flux_div_rel}
and \ref{eq:disc_y_mom_flux_div_rel}.  As we will show through the tests
described in \S3.4, the Crank-Nicholson time differencing of the shearing
box source terms used in these equations conserves the energy integral
associated with epicyclic motion exactly.

To implement these difference equations in a computer code additional
algorithmic steps are required.  For example, the reconstruction of
the left- and right-interface states in the $x-$direction (see \S4.2
of SGTHS) which is needed to compute the time-and area averaged fluxes
of the conserved variables at the $x-$interfaces using a Riemann solver
requires the addition of shearing box source terms.  In particular, before
the left- and right-states in the primitive variables at $x-$interfaces
are converted back to the conserved variables (${\bf q}_{L,i-1/2}$ and
${\bf q}_{R,i-1/2}$ in the notation of SGTHS) at the end of the first-,
second-, or third-order reconstruction algorithms describe in \S4.2 of
SGTHS, the shearing box source terms for the velocity components $v_x$
and $v_y$ must be added for $\delta t/2$.  We have found that simple
forward Euler time-differencing is adequate for this step.

Similarly, the directionally unsplit corner transport upwind (CTU)
integrator used in Athena (see \S5.1 and \S6.1 in SGTHS) uses transverse
flux gradients to correct the interface states in multidimensions.  For
the flux gradients in the $x-$direction (added to the $y-$interface states
in 2D, and the $y-$ and $z-$interface states in 3D), the appropriate
shearing box source terms must be added to $x-$ and $y-$components of
the momentum for $\delta t/2$.  Again, we have found that simple forward
Euler differencing is adequate for this step.

Finally, in order to compute the cell-centered reference electric
field at the half time step $\E^{r,n+1/2}_{i,j,k}$ needed for the
CT algorithm (computed in step 5 of the 2D CTU algorithm, or step 6 in the 
3D CTU algorithm), the shearing box source terms 
with forward Euler discretization must be added in
the calculation of the velocity components at the half time step.

In summary, our algorithm for the momentum equation update in the shearing
box approximation as implemented in Athena
consists of the following modifications to the
CTU+CT algorithm described in SGTHS:
\begin{enumerate}
\item Add shearing box source terms to the left- and right-states for
the velocity at $x-$interfaces only during reconstruction step, before
converting the reconstructed primitive variables to conserved variables.
\item Add shearing box source terms to the $x-$ and $y-$components of the
momentum when a transverse flux gradient in
$x-$direction is applied to $y-$ and $z-$interface states as part of CTU
integrator.
\item Add shearing box source terms to velocity when computing the cell-centered
reference electric field at the half time step needed by the CT algorithm
(see step 5 in \S 4.2 of Gardiner \& Stone 2008).
\item Use equations \ref{eq:m_cons_evol_2} and \ref{eq:rhovy_cons_evol_3}
to evolve the momentum in the final conservative update, where the
divergence of the fluxes of the momentum fluctuations come from equations 
\ref{eq:disc_x_mom_flux_div_rel} and \ref{eq:disc_y_mom_flux_div_rel}.
\end{enumerate}
As we show with tests, the main advantage of this algorithm is that it
conserves the energy integral for epicyclic motion exactly.

\subsection{Source Terms in the Energy Equation}

Next we consider the finite-volume discretization of the energy equation
\ref{eq:cons_energy} including the shearing box source terms.
The key to this discretization is
to define an effective potential for the shearing box
\begin{equation}
\Phi_{SB} = -q\Omega_{0}^{2}x^{2} + \frac{1}{2}\Omega_{0}^{2}z^{2}
\label{eq:effective-pot}
\end{equation}
so that the source terms in the energy equation \ref{eq:cons_energy} can
be written as
\begin{equation}
\frac{\partial E}{\partial t} +
\nabla\cdot \left[E{\bf v} + {\sf T} \cdot {\bf v}\right]  =  
\rho{\bf v} \cdot \nabla \Phi_{SB}
\label{eq:energy+potential}
\end{equation}
When written in this form, it is clear that the source term represents
the rate of change in the gravitational potential energy per unit volume.

We seek a discretization of this term such that when integrated over volume,
we recover the rate of change of energy due to the work done at the boundaries.
This suggests that for each computational cell, we discretize the source term
by the difference of the work done at the edges of the cell.
Ignoring for the moment the vertical component of gravity, so that the effective
potential depends only on $x$, the appropriate
finite-volume discretization of equation \ref{eq:energy+potential} is
\begin{eqnarray}
E_{i,j,k}^{n+1} &=& E_{i,j,k}^n - \delta t
\overline{\left( {\bf \nabla} \cdot {\bf F}_{E} \right)}_{i,j,k}
\nonumber \\
&-&\frac{\delta t}{2} \left( 
(F_\rho)_{x,i+1/2,j,k} \frac{\Phi_{SB,i+1/2,j,k}-\Phi_{SB,i,j,k}}{\delta x /2}
\right. \nonumber \\
&+&
\left. 
(F_\rho)_{x,i-1/2,j,k} \frac{\Phi_{SB,i,j,k}-\Phi_{SB,i-1/2,j,k}}{\delta x /2}
\right)
\label{eq:E_cons_evol}
\end{eqnarray}
where $(F_\rho)_{x,i+1/2,j,k}$ are the mass fluxes in the $x-$direction
returned by the Riemann solver at the $x-$interfaces.
It is straightforward to confirm that when integrated over volume, the
source term reduces to the net mass flux times the difference in the
gravitational potential across the domain.
Thus, the only route via which the total energy in the domain can change
is through work done at the boundaries.  Within the domain energy can
be exchanged between its kinetic, magnetic, and thermal forms, but it
cannot be lost as truncation error.  Even though we evolve an energy
variable that does not contain the gravitational potential energy, we use
a discretization that conserves to machine precision the {\em total} energy,
that is $E + \rho \Phi_{SB}$.

In addition to the use of the update equation \ref{eq:E_cons_evol},
one other algorithmic step is required to include the shearing box
source terms in the energy equation.  This step is to include source
terms to the transverse flux gradient corrections to the left- and
right-interface states for $E$ in the multidimensional CTU algorithm.
To be more precise, when the flux gradients in the $x-$direction
are added to the $y-$interface states of $E$ in 2D, and the $y-$ and
$z-$interface states of $E$ in 3D, the appropriate shearing box source
terms must included for $\delta t/2$.  Again, we have found that
simple forward Euler differencing is adequate for this step.  Note
that since the reconstruction step uses the primitive ($p$) rather
than conserved ($E$) variables, no additional source terms are
required for the evolution of the energy equation in the reconstruction
step.

Thus, the extensions to the algorithms for energy equation update in the
shearing box approximation consists of the following modifications:
\begin{enumerate}
\item Add shearing box source terms to $E$ when a transverse flux gradient
in $x-$direction is applied to $y-$ and $z-$interface states of $E$
as part of the CTU integrator.
\item Use equation \ref{eq:E_cons_evol}
to evolve $E$ in the final conservative update.
\end{enumerate}
As shown in Gardiner \& Stone (2005b) in studies of the MRI in the
shearing box with Athena, with this algorithm the rate of change of $E$
is equal to the work done at the boundaries to machine precision.

\subsection{Including Vertical Gravity}

To simplify the discussion in the previous subsections, we considered
only the two components of the momentum equation in the orbital ($x-y$)
plane, and the source term associated with the vertical component of
gravity in the energy equation was ignored.  We have shown that care is
needed in the discretization of the source terms in these equations in
order to correctly capture the dynamics of epicyclic motion.  However,
for the discretization of the gravitational acceleration term in the
vertical ($z-$) component of the momentum equation, and the work term
associated with the vertical component of gravity in the energy equation,
we have found less complicated methods are adequate.

In Athena, the vertical gravity terms in {\em both} the vertical component
of the momentum and the energy equation are added by differencing the
effective potential equation \ref{eq:effective-pot}.  For the vertical
component of the momentum equation, several algorithmic steps are
required:
\begin{enumerate}
\item At the end of the reconstruction step, vertical acceleration
for $\delta t/2$ is added to the left- and right-states of $v_{z}$
at $z-$interfaces using the vertical gradient of the potential.
\item When the transverse flux corrections in the $z-$direction are applied
to the $x-$ and $y-$interface states as part of the CTU unsplit
integrator, vertical acceleration for $\delta t/2$ is added to $\rho v_{z}$
using the gradient of the potential.
\item When computing the cell-centered
reference electric field at the half time step needed by the CT algorithm 
(see step 5 in \S 4.2 of Gardiner \& Stone 2008),
vertical acceleration for $\delta t/2$ is added to $v_{z}$ using the 
gradient of the potential.
\item In the final conservative update, the vertical force for $\delta t$ is
added to $\rho v_{z}$ using the 
gradient of the potential and the density evaluated at the half time step,
$(\rho^{n}_{i,j,k} + \rho^{n+1}_{i,j,k})/2$. 
\end{enumerate}

For the energy equation, the final conservative update equation \ref{eq:E_cons_evol} must be extended to include a term $S_{E,z}$, where
\begin{eqnarray}
S_{E,z} &=& -\frac{\delta t}{2} \left( 
(F_\rho)_{z,i,j,k+1/2} \frac{\Phi_{SB,i,j,k+1/2}-\Phi_{SB,i,j,k}}{\delta z /2}
\right. \nonumber \\
&+&
\left. 
(F_\rho)_{z,i,j,k-1/2} \frac{\Phi_{SB,i,j,k}-\Phi_{SB,i,j,k-1/2}}{\delta z /2}
\right)
\label{eq:E_cons_evol_zterm}
\end{eqnarray}
This leads to the following algorithmic steps:
\begin{enumerate}
\item Add shearing box source terms to $E$ when a transverse flux gradient
in $z-$direction is applied to $x-$ and $y-$interface states of $E$
as part of the CTU integrator.
\item Use equation \ref{eq:E_cons_evol}, including the term given by
equation \ref{eq:E_cons_evol_zterm} on the RHS,
to evolve $E$ in the final conservative update.
\end{enumerate}
Inclusion of the source terms for vertical gravity, along with appropriate
boundary conditions in the vertical direction, allow studies
of the MRI in stratified disks (e.g. Stone et al. 1996) to be continued
using Athena (Davis et al. 2010).

\subsection{Tests of the Source Terms}

There are a variety of axisymmetric ($\partial / \partial y \equiv 0$)
solutions to the MHD equations in the shearing box that serve as useful
tests of the implementation of the source terms.  Non-axisymmetric
solutions required a more sophisticated treatment of the boundary
conditions, as described in the next section.  All of the tests in this
section use a periodic domain in the $(x-z)$ plane which spans $-L_{x}/2
\leq x \leq L_{x}/2$ and $-L_{z}/2 \leq z \leq L_{z}/2$. The orbital
frequency $\Omega_{0}=10^{-3}$, the shear parameter $q=3/2$, and an
isothermal equation of state is used with sound speed $C_s = 10^{-3}$.
Unless otherwise stated, the initial conditions consist of a uniform
density medium with $\rho_{0} = 1$, pressure $p=\rho_{0}C_{s}^{2}$,
and orbital velocity $v_{y}(x,0) = -q\Omega_{0} x$.  Third-order
reconstruction, the HLLC (for hydrodynamics) or HLLD (for MHD) Riemann
solver, and the CTU+CT unsplit integrator are used in all the tests.

A good first test is the evolution of epicyclic oscillations.  We choose
the initial radial velocity $v_{x}=0.1C_s$, where $C_s=10^{-3}$,
in domains of size $L_{x}=L_{z}=1$, $10$, and $50$, and evolve the
flow for thousands of orbits using a grid of $64^{2}$.  We vary
the size of the domain while keeping the resolution constant in order
to study the effect of very large timesteps on the accuracy of the
integration algorithm.  Figure 1 shows the oscillations in $v_{x}$ over
the first 20 orbits in each run.  In each case, we observe epicyclic
motion with constant amplitude: the energy integral in epicyclic
motion is conserved exactly.  Note there is a small dispersion error
in the largest domain.  However, in this case the time step is so large
that there are only 14 timesteps
per orbit.  Thus, the small dispersion error ($\lapprox 2$\%) observed in
this case (due to the Crank-Nicholson differencing) is to be expected.
Increasing the resolution, and therefore decreasing the timestep, makes
this error converge at second order.  The key result is that even with
large timesteps, the amplitude of oscillations is constant.  However,
we emphasize that this result is relevant only for hydrodynamics, since
with weak magnetic fields epicyclic motion is unstable.

A more quantitative test is provided by the propagation of nonlinear
axisymmetric density waves in the shearing box (Fromang \& Papaloizou
2007).  The profile of such waves is given by the solution to two
ordinary differential equations; for this test we use a numerical solution
(kindly provided by S. Fromang) on a grid of 40 points in a domain with
$L_{x}=9.593$ with a wave with amplitude $\rho_{\rm max}/\rho_{0} = 1.05$.
Figure 2 shows the error in the density, defined as $\delta \rho(x,t) =
(\rho(x,t)-\rho(x,0))/\rho(x,0)$ after $t=10$, 50, and 100 wave periods
$T$, where $T=5254.31$.  Note the error is smooth (there are no oscillations
indicative of dispersion error), and never larger than 1.5\%.  We also
plot the time evolution of the error in the wave amplitude, defined as
$(\epsilon - \epsilon_0)/\epsilon_0$, where $\epsilon = {\rm max}(\rho)
- \rho_0$.  After 100 wave periods, we find the amplitude has decreased
by only 3\%, which is a measure of the low numerical diffusion rate
in Athena.  These results compare very favorably to the test results
presented in Fromang \& Papaloizou (2007), which were computed at a much
higher resolution.

Finally, another useful test is the nonlinear evolution of
axisymmetric modes of the MRI with no net flux.  We use a domain of
size $L_{x}=L_{z}=1$, and an initially vertical field with ${\bf B} =
(0,0,(2p_{0}/\beta)^{1/2}\sin{2\pi x/Lx})$, where $\beta=4000$.  This test
is identical to run S1c in Hawley \& Balbus (1992).  In figure 3, we
show images of the azimuthal velocity fluctuations $v_y + q\Omega_{0}x$
at various times in the evolution.  The growth and saturation of the MRI
is evident, with transition to two-dimensional MHD turbulence at late
times.  In two dimensions, turbulence cannot be sustained by the MRI,
and the magnetic energy eventually decays.  The rate of decay depends
on numerical diffusion, which in turn depends on the grid resolution
and numerical algorithm.  Previously, we have used the rate of decay of
the poloidal magnetic energy after saturation to compare the numerical
dissipation rates of ZEUS and Athena (Stone \& Gardiner 2005; Stone 2009),
to show that Athena gives lower decay rates at the same grid resolution.
Figure 4 compares the time evolution of the energy associated with
the radial component of the magnetic field $B_{x}^{2}/2$ for various
resolutions and both second- and third-order reconstruction.  At every
resolution the third-order reconstruction is measurably more accurate,
in the sense that the rate of decay of the magnetic energy is decreased.
Perhaps the best indicator of the accuracy of the methods is the evolution
at very early times, while the amplitude of the MRI is very small.
The short period of decay (due to the expansion of regions with the
highest magnetic pressure), followed by exponential growth over 10 orders
of magnitude in energy from the initial (very small) amplitude indicates
Athena has a very good dispersion relation for slow MHD waves (which
was also shown in the Stone et al. 2008 linear wave convergence tests).

\section{Shearing Box Boundary Conditions}

As first discussed in HGB, non-axisymmetric
solutions in the shearing box require special boundary conditions that
offset the solutions by the distance the radial edges of the
domain have been displaced by the background shear.  Mathematically,
the shearing box boundary conditions can be expressed as
(HGB; Gressel \& Ziegler 2007):
\beq
f(x,y,z) \longmapsto f(x\pm L_{x},y \mp wt,z)
\eeq
where $f = (\rho, \rho v_{x}, \rho v_{z}, {\bf B})$, and
\beq
\rho v_{y}(x,y,z) \longmapsto \rho v_{y}(x\pm L_{x},y \mp wt,z)
\mp \rho w
\label{eq:vy-remap}
\eeq
\beq
E(x,y,z) \longmapsto E(x\pm L_{x},y \mp wt,z)
\mp \rho v_{y} w + \rho w^{2}/2
\label{eq:ener-remap}
\eeq
where $w=q\Omega_{0}L_{x}$ is the difference in the orbital velocity
across the domain.  Although a variety of authors have already
described the implementation of the shearing box boundary conditions in
various codes (e.g. HGB; Gressel \& Ziegler 2007), the details of the
implementation can be important, thus we describe our methods below.

In Athena, we implement the shearing box boundary conditions by
first applying periodic boundary conditions in the radial direction,
(with the appropriate shift in the $y-$momentum and energy given by
equations \ref{eq:vy-remap} and \ref{eq:ener-remap}), and then
use a conservative remap of all 
quantities in the ghost cells in the
$y-$direction by a displacement $wt$, the distance the boundaries
have sheared in time $t$.  This distance in general is not an integer
number of cells.  To remap the solution a fractional portion of a cell
we use the same higher order reconstruction functions as used in the
integrator to compute the amount of each conserved variable to assign to
the remapped cells.  For the magnetic field, we remap each component
in the ghost zones directly, rather than evolving the field in the
ghost zones using a remap of the emfs (e.g. Gressel \& Ziegler 2007).
This procedure does not guarantee the divergence free constraint is
maintained {\em in the ghost zones}.  However, since these cells are
never used for anything more than interpolation, we have not found this
to be a problem.

As pointed out by Gressel \& Ziegler (2007), the shearing box boundary
conditions can destroy conservation, because the integral of the fluxes of
the conserved quantities over the two radial faces may not be identical
due to the remap.  In particular, if the integral in the $y-$direction
of the azimuthal component of the EMF at each radial face is not equal,
then the net vertical flux in the domain is not conserved.  Since the
dynamics of the MRI is sensitive to the amplitude of the net vertical
flux, this can represent a serious problem.  To circumvent this issue,
we remap the azimuthal component of the emf at each radial face, and use
the arithmetic average of $\E_{y}$ computed for each grid cell at each
radial face, and the remapped value of $\E_{y}$ from the corresponding
grid cell on the opposite face, to update the magnetic field in the
cells next to the boundary.  Since the remap conserves the integral of
$\E_{y}$ in the $y-$direction, the integral of the averaged $\E_{y}$
used to update the field on each boundary will be identical, and so this
procedure conserves the net flux in the vertical direction to machine
precision.  In principle, this same procedure could be used for the fluxes
of all the conserved quantities, to restore conservation in the domain.
In particular, it can also be used to remap $\E_{z}$, so that the volume
averaged azimuthal flux of the magnetic field is conserved to machine
precision.  In practice, we find that with the orbital advection algorithm
described in \S5, the hydrodynamic variables are conserved accurately
even without this extra step, so that we generally only correct $\E_{y}$
in this fashion.

In the same way that the net vertical flux in the domain is not
conserved without a special treatment of the remap of  $\E_{y}$,
the net toroidal flux is not conserved without a special treatment of
the remap of $\E_{z}$.  We have run a variety of simulations of the
MRI in three dimensions, and find the net toroidal flux introduced in
this way is very small: the magnetic energy associated with this flux
shows random fluctuations with peak amplitude of around $10^{-9}$ of
the thermal pressure for a zero-net-vertical flux simulation with an
initial $\beta=4000$.  However, we have found that matching the order
of the reconstruction used in the ghost cells with the order used in
the integrator, that is using third-order reconstruction in the
ghost cells when third-order reconstruction is used throughout the
rest of the domain, is important to keep this anomalous flux small.
Using second-order reconstruction in the ghost zones produced a mean
toroidal flux that is still very small, but 10 times larger than when
third-order reconstruction is used in the ghost zones.

Efficient use of a large number ($\gapprox 10^{3}$) of processors
on modern parallel computers requires domain decomposition of the
grid in all three coordinate directions.  However, parallelization
of the shearing box boundary conditions can require complex
book-keeping if domain decomposition in the azimuthal direction is
allowed, because the data in the ghost zones for each processor on
the radial boundaries can require communication with up to 3
processors on the opposite boundary, depending on the configuration
of the domains as they shear.  To simplify the implementation, we
first apply periodic boundary conditions (which are easily parallelized)
to pass ghost cells between processors in the radial direction,
followed by a remap in the azimuthal direction.  The latter is
easily parallelized using a cyclic shift operation across those
processors that store neighboring grid patches in the azimuthal
direction.  The cyclic shift is also easily parallelized, and always
occurs between the same set of processors along the boundary.  The
disadvantage of this method is that it adds extra MPI calls, however
this is more than offset by the reduction in complexity of the code,
and the fact that the resulting communication pattern is more regular
and therefore in most circumstances likely to be more efficient.

\section{Orbital Advection}

Orbital advection methods (Masset 2000; JGG; Johansen et al. 2009)
exploit the fact that the background orbital motion ${\bf
v}_{K}$ is time-independent and varies only in radius (the $x-$coordinate in
the shearing box).  Thus, by decomposing the velocity into two parts
\beq
{\bf v} = {\bf v}_{K} + {\bf v}^{\prime}
\eeq
where the components of the velocity fluctuation vector ${\bf v}^{\prime}$
are
\begin{equation}
\left[ \begin{array}{c}
  v_{x}^{\prime} \\
  v_{y}^{\prime} \\
  v_{z}^{\prime} \end{array} \right] =
\left[ \begin{array}{c}
  v_{x} \\
  v_{y} + q\Omega_{0}x \\
  v_{z} \end{array} \right]
\label{eq:vel-fluctuations}
\end{equation}
the equations of MHD in
the shearing box (equations \ref{eq:cons_mass} through \ref{eq:induction})
can be rewritten as
\begin{eqnarray}
\frac{\partial \rho}{\partial t} + v_{K}\frac{\partial \rho}{\partial y}
+{\bf\nabla\cdot} [\rho{\bf v}^{\prime}]& =& 0,
\label{eq:fargo_cons_mass} \\
\frac{\partial \rho {\bf v}^{\prime}}{\partial t} +
v_{K} \frac{\partial \rho {\bf v}^{\prime}}{\partial y} +
{\bf\nabla\cdot} \left[\rho{\bf v}^{\prime}{\bf v}^{\prime} + {\sf T}\right] &=&
2\Omega_{0} (\rho v_{y}^{\prime}){\bf \hat{i}}
+ (q-2)\Omega_{0} (\rho v_{x}^{\prime}){\bf \hat{j}}
- \rho \Omega_{0}^{2}z{\bf \hat{k}}
\label{eq:fargo_cons_momentum} \\
\frac{\partial E^{\prime}}{\partial t} + 
v_{K} \frac{\partial E^{\prime}}{\partial y} +
\nabla\cdot \left[E^{\prime}{\bf v}^{\prime} + {\sf T} \cdot {\bf v}^{\prime}\right] &=&
\rho{\bf v}^{\prime} \cdot {\bf \nabla}\Phi^{\prime}_{\rm SB}
+ (B_{x}B_{y} - \rho v_{x}^{\prime}v_{y}^{\prime})\frac{\partial v_{K}}{\partial x}
\label{eq:fargo_cons_energy} \\
\frac{\partial {\bf B}}{\partial t} -
{\bf\nabla} \times \left({\bf v}_{K} \times {\bf B}\right) & - &
{\bf\nabla} \times \left({\bf v}^{\prime} \times {\bf B}\right) = 0, 
\label{eq:fargo_induction}
\end{eqnarray}
where the total energy $E^{\prime}$ contains the kinetic energy in the
velocity fluctuations (but not the background shear flow)
\begin{equation} 
E^{\prime} = \frac{p}{\gamma -1} + 
\frac{1}{2}\rho ({\bf v}^{\prime}\cdot{\bf v}^{\prime}) + \frac{B^{2}}{2},
\label{eq:fargo_totalenergy}
\end{equation}
and the effective potential for the shearing box in the orbiting frame 
contains only the vertical gravity
\begin{equation}
\Phi^{\prime}_{SB} = \frac{1}{2}\Omega_{0}^{2}z^{2}
\label{eq:fargo_effective-pot}
\end{equation}
Note the second term on the LHS in each of equations \ref{eq:fargo_cons_mass}
through \ref{eq:fargo_induction} is a linear advection term with
characteristic speed $v_{K}$.  This suggests a numerical algorithm based
on splitting these equations into two systems, the first containing only
the linear advection operators
\begin{eqnarray}
\frac{\partial \rho}{\partial t} + v_{K}\frac{\partial \rho}{\partial y}
 & = & 0,
\label{eq:advect_cons_mass} \\
\frac{\partial \rho {\bf v}^{\prime}}{\partial t} + v_{K}
\frac{\partial \rho{\bf v}^{\prime}}{\partial y} & = & 0
\label{eq:advect_cons_momentum} \\
\frac{\partial E^{\prime}}{\partial t} + v_{K}
\frac{\partial E^{\prime}}{\partial y} & = & 
(B_{x}B_{y}-\rho v_{x}^{\prime}v_{y}^{\prime})\frac{\partial v_{K}}{\partial x}
\label{eq:advect_cons_energy} \\
\frac{\partial {\bf B}}{\partial t} -
{\bf\nabla} \times \left({\bf v}_{K} \times {\bf B}\right) & = & 0
~.
\label{eq:advect_induction}
\end{eqnarray}
Note the extra terms on the RHS in the energy equation.  These terms cannot
be written as flux gradients, and must be treated as source terms.  We
discuss them in more detail below.

The remaining system of equations are the usual equations of MHD, written
in terms of the velocity fluctuations ${\bf v}^{\prime}$ rather than
the velocity ${\bf v}$, that is
\begin{eqnarray}
\frac{\partial \rho}{\partial t} +
{\bf\nabla\cdot} [\rho{\bf v}^{\prime}] & = & 0,
\label{eq:fluct_cons_mass} \\
\frac{\partial \rho {\bf v}^{\prime}}{\partial t} +
{\bf\nabla\cdot} \left[\rho{\bf v}^{\prime}{\bf v}^{\prime} + {\sf T}\right] & = &
2\Omega_{0} (\rho v_{y}^{\prime}){\bf \hat{i}}
+ (q-2)\Omega_{0} (\rho v_{x}^{\prime}){\bf \hat{j}}
- \rho \Omega_{0}^{2}z{\bf \hat{k}}
\label{eq:fluct_cons_momentum} \\
\frac{\partial E^{\prime}}{\partial t} + 
\nabla\cdot \left[E^{\prime}{\bf v}^{\prime} + {\sf T} \cdot {\bf v}^{\prime}\right] & = &
\rho{\bf v}^{\prime} \cdot {\bf \nabla}\Phi^{\prime}_{\rm SB},
\label{eq:fluct_cons_energy} \\
\frac{\partial {\bf B}}{\partial t} -
{\bf\nabla} \times \left({\bf v}^{\prime} \times {\bf B}\right) & = & 0, 
\label{eq:fluct_induction}
\end{eqnarray}
Note that since these equations written using $\rho v_{y}^{\prime}$
and $E^{\prime}$ as the conserved variables, instead of $\rho v_{y}$
and $E$, the shearing box source terms are modified compared to the
original system equations \ref{eq:cons_mass} through \ref{eq:induction}.

Developing numerical algorithms to solve these two systems of equations is 
straightforward.  In particular, since the second system equations
\ref{eq:fluct_cons_mass} through \ref{eq:fluct_induction} are just the
usual equations of MHD, but written in terms of a vector of
conserved variables ${\bf q} = (\rho,\rho{\bf v}^{\prime},E^{\prime},{\bf B})$,
the standard CTU+CT algorithm in Athena can be used to integrate these
equations,
with the appropriate Crank-Nicholson time differencing of the shearing box
source terms in the momentum fluctuation equation, and the gradient of
the effective gravitational potential in the energy equation.  These
source terms are in fact simpler than in the original system, and the
techniques described in \S3 apply directly.  Since the variables being
updated are the momentum fluctuations themselves, the source terms added to
the reconstruction step and transverse flux gradient corrections (see \S3.2)
are modified appropriately.  Moreover, since the Riemann solver now returns
the fluxes of the momentum fluctuations directly, the conversions in equations
\ref{eq:disc_x_mom_flux_div_rel} and \ref{eq:disc_y_mom_flux_div_rel}
are no longer necessary.  Note that the CFL stability
condition for this system is based only on the amplitude of the
velocity fluctuations, that is
\begin{equation}
\delta t = C_{\circ} \min \left( 
\frac{\delta x}{(\vert v^{\prime}_{x} \vert + C_{fx})_{i,j,k}},
\frac{\delta y}{(\vert v^{\prime}_{y} \vert + C_{fy})_{i,j,k}},
\frac{\delta z}{(\vert v^{\prime}_{z} \vert + C_{fz})_{i,j,k}} \right)
\end{equation}
where $C_{\circ} \le 1$ is the CFL number, and $C_{fx}$, $C_{fx}$,
$C_{fx}$ are the fast magnetosonic waves speeds in the $x-$, $y-$,
and $z-$directions respectively, and the minimum is taken over all
grid cells.  Since the background orbital flow becomes supersonic for
$\vert x \vert > H/q$, where $H=C_{s}/\Omega_{0}$ is the scale height
in the disk, the time step can be much larger using orbital advection
in domains which span roughly $H$ or larger in the radial direction.
This is the primary advantage of the method.

Since the first system of equations (\ref{eq:advect_cons_mass} through \ref{eq:advect_induction}) are linear advection operators in one dimension, 
the numerical algorithms to integrate these equations are particularly simple.
The finite volume
discretization of the first three of these equations can be written as
\beq
U_{i,j,k}^{n+1} = U_{i,j,k}^{n} - (\delta t / \delta y)
(\overline{f}_{i,j+1/2,k} - \overline{f}_{i,j-1/2,k})
\label{eq:fargo-update}
\eeq
where $U$ represents each of the conserved quantities, and
$\overline{f}_{i,j+1/2,k}$ 
the upwind fluxes of these quantities 
at cell interfaces in the 
$y-$direction.  These fluxes 
are simply the integral of $U$
over the domain $v_{K}t$ upstream of the appropriate interface, for example
\beq
\overline{f}_{i,j-1/2,k} 
 = \int_{(y_{j-1/2})-v_{K}\delta t}^{y_{j-1/2}} U^{n}(x_{i},y,z_{k})dy
\label{eq:fargo-flux}
\eeq
Numerically, this integral is converted into a finite sum over all the grid
cells upstream of the interface in the $y-$direction.  In general,
the domain of dependence 
$v_{K}\delta t$ will span a non-integer number of cells.  For those cells
entirely contained in the domain, the 
volume-averaged value
of $U$ is added to the sum.  For the one cell which is only partially
contained within the integral, then the higher-order reconstruction
functions described in \S4 of SGTHS are used to evaluate the fraction
of the conserved variable that contributes to the integral.  Note there
is no time step constraint for stability of this algorithm.  The domain
of dependence can be arbitrarily large, and span any number of grid cells.
The step simply represents a conservative remap of the solution in the 
$y-$direction by a distance $v_{K}\delta t$.

There are two aspects to the orbital integration algorithm that warrant
some discussion.  The first is the integration of the induction equation.
Note we have written equation \ref{eq:advect_induction}
in a form that suggests the use of the CT algorithm with an effective
emf given by ${\bf \E} = -{\bf v}_{K} \times {\bf B}$.  The
components of this emf are
\beq
(\E_{x}, \E_{y}, \E_{z}) = (-v_{K}B_{z}, 0, v_{K}B_{x})
\eeq
The centering of the components of the emf used in Athena are shown
in figure 1 in SGTHS.  The discrete form of the CT update for each
component of the magnetic field is given by equations 16 through 18
in SGTHS.  The CT algorithm for orbital advection simply
requires the calculation of the effective emf by integration of each
component of the electric field ${\bf \E}$ upstream of the appropriate
cell edges distance $v_{K}\delta t$, that is
\beq
\E^{n+1/2}_{x,i,j-1/2,k-1/2} = - \int_{(y_{j-1/2})-v_{K}\delta t}^{y_{j-1/2}}
 v_{K}B_{z}^{n}(x_{i},y,z_{k-1/2})dy
\label{eq:fargo-ex-flux}
\eeq
\beq
\E^{n+1/2}_{z,i-1/2,j-1/2,k} = \int_{(y_{j-1/2})-v_{K}\delta t}^{y_{j-1/2}}
 v_{K}B_{x}^{n}(x_{i-1/2},y,z_{k})dy
\label{eq:fargo-ez-flux}
\eeq
By using a CT discretization of
equation \ref{eq:advect_induction} to evolve the magnetic field in the
orbital advection step, we
preserve the divergence-free constraint to machine round-off during the
orbital advection step.  Moreover, note that there are no source terms
required in equation \ref{eq:advect_induction}.  The growth or decay of
$B_{y}$ due to the shear is naturally captured in the CT difference
formulae, avoiding any divergence-free interpolation as required in the
algorithm of JGG.

The second aspect of the orbital advection step that requires
further discussion is the integration of the energy equation
\ref{eq:advect_cons_energy}.  Note the source terms that appear on
the RHS, which represent the work done by Reynolds and magnetic stress
due to the radial shear of the orbital
motion.  The finite-volume discretization of equation
\ref{eq:advect_cons_energy} requires a time- and volume-averaged
approximation for the source terms.  In the Lagrangian frame (comoving
with the fluid during the remap), all quantities except $B_{y}$ are
constant, and the time evolution of $B_{y}$ in the Lagrangian frame
is particularly simple, $B_{y}(t+\delta t) = B_{y}(t) + B_{x}(\partial
v_{K}/\partial x)\delta t$, where $\partial
v_{K}/\partial x = -q\Omega_{0}$  This suggest a ``Lagrangian-step-plus-remap"
algorithm for the energy equation.  In the Lagrangian step, the total
energy is updated using the time-averaged source terms 
\begin{eqnarray}
\hat{E}^{\prime}_{i,j,k} &=& E^{\prime}_{i,j,k} 
- q\Omega_{0}\delta t B_{x,i,j,k}(B_{y,i,j,k}(0)
- B_{x,i,j,k}q\Omega_{0}\delta t/2) \nonumber \\
 &+&  q\Omega_{0}\delta t \rho  v^{\prime}_{x,i,j,k}v^{\prime}_{y,i,j,k}
\end{eqnarray}
where $B_{x,i,j,k}$ and $B_{y,i,j,k}(0)$ are the volume-averaged,
cell-centered components of the magnetic field at the start of
the orbital advection step.  The remap step then uses equation
\ref{eq:fargo-update} applied to the total energy with the source
terms added, $\hat{E}^{\prime}_{i,j,k}$, to complete the update of the
total energy.  Comparisons of calculations of the MRI with an adiabatic
equation of state, both with and without orbital advection (see next
section), show that this algorithm for the energy source terms works well.
Note that the only way in which the total volume averaged energy in the
domain can change is via a non-zero total stress at the radial
boundries.  Additional tests have shown our method conserves the
total energy if the stress at the boundaries is zero.

\subsection{Tests of Orbital Advection}

To test our orbital advection algorithm, we have run calculations in
both two dimensions (in the $x-y$ plane), and full three dimensions.
Our calculations span $-L_{x}/2 \leq x \leq L_{x}/2$,
$-L_{y}/2 \leq x \leq L_{y}/2$, and $-L_{z}/2 \leq z \leq L_{z}/2$.
We set the orbital frequency $\Omega_{0}=10^{-3}$, the shear parameter
$q=3/2$, and unless otherwise stated
adopt an isothermal equation of state with sound speed $C_s$
(we give the value of $C_s$ used for each test below).  As before, we
set the initial density $\rho_{0} = 1$ and orbital velocity $v_{y,0}
= -q\Omega_{0} x$, and we use third-order reconstruction, the HLLC (for
hydrodynamics) or HLLD (for MHD) Riemann solver, and the CTU+CT unsplit
integrator for all the tests.

Our first test of orbital advection is the evolution of
a hydrodynamic shearing wave (Johnson \& Gammie 2005, Balbus \& Hawley
2006; Shen et al. 2006).  We use a domain of size $L_{x}=L_{y}=1$,
an isothermal equation of state with sound speed $C_{s}=1.29 \times
10^{-3}$, and an initial wavevector $2\pi {\bf k}_{0}/L =(-8, 2)$.
We expect the maximum amplification of the wave amplitude to be 17, at
time $t\Omega_{0}=8/3$.  Figure 5 compares the amplitude of the kinetic
energy associated with the $x-$component of the velocity both with
and without the orbital advection algorithm at various resolutions.
In both cases, the evolution is accurately captured provided there
are at least 8 grid points per wavelength in the initial conditions.
Most importantly, there is virtually no aliasing in the solutions at
late times.  In both cases, the kinetic energy decays over 5 orders of
magnitude, with no periods of sustained growth after the maximum.

Our first full MHD test of orbital advection is the advection of a weak 
($\beta = 2\times 10^{6}$) field loop.
This test has been useful for developing the
CT algorithm implemented in Athena (Gardiner \& Stone 2005a; 2008;
Stone et al. 2008).
We use a domain of size $L_{x}=3$ and $L_{y}=8$, an isothermal equation
of state with $C_{s}=10^{-3}$, and initialize the field loop following the
method described in GS05 centered at the origin.  To make the problem
more complex, we introduce a uniform radial velocity $v_{x}=C_{s}$,
which causes the loop to execute epicyclic oscillation as it is sheared.
The amplitude of these oscillations is large enough so that the outer one
quarter of the loop crosses the radial boundary at the extrema of
the motion.  This tests whether our implementation of the shearing box
boundary conditions described in \S4 can maintain the integrity of the
loop.  Figure 6 shows plots of the field lines at four times in the
evolution.  Note there are no indications of the stripes which
appear if the EMFs are computed incorrectly (see figures
2 and 3 in GS05), confirming
the upwind CT method developed by GS05 to compute the emfs also works well
with orbital advection.  In addition, there are no features in the loop
associated with the boundaries, indicating our implementation of the
shearing box boundary conditions is accurate.

Another sensitive MHD test recently introduced by JGG is the evolution
of a compressible shearing wave (Johnson 2007).  We have repeated the
test shown in figure 11 of JGG, and compared the resulting solution to a
numerical integration of the ODEs that describe the analytic solution to
the problem kindly provided to us by B. Johnson.  The calculation uses
a domain of size $L_{x}=L_{y}=L_{z}=L=0.5$, and a numerical resolution
of $N \times N/2 \times N/2$, where $N=16$, 32, and 64 (these values are
chosen to match those used by JGG).  An isothermal
equation of state with $C_{s}=1$ is adopted, and for this test we use
$\Omega_{0}=1$.  Although the details of how to initialize such tests
are given in the Addendum to JGG, to be specific we give the initial
conditions we actually used in Athena for this test, namely density
$\rho=1+5.48082\times 10^{-6}\cos{({\bf k}_{0}\cdot{\bf x})}$, velocity
fluctuations ${\bf v}^{\prime}=(-4.5856\times 10^{-6},2.29279\times
10^{-6},2.29279\times 10^{-6})\cos{({\bf k}_{0}\cdot{\bf x})}$, and
magnetic field ${\bf B}/\sqrt{4\pi} =(5.48082\times 10^{-7},1.0962\times
10^{-6},0)\cos{({\bf k}_{0}\cdot{\bf x})} + (0.1,0.2,0)$, where the
wavevector ${\bf k}_{0}$ has components $(-2,1,1)(2\pi/L)$.
Figure 7 shows the time evolution of the volume-averaged
perturbations in the real part of the
azimuthal component of the magnetic field, defined as
\beq
\delta B_{y} = \int_{V} 2(B_{y} - \hat{B}_{y})\cos{({\bf k}(t)\cdot {\bf x})}
\eeq
where $B_{y}$ is the numerical solution computed by Athena, $\hat{B}_{y}
= 0.2-0.15\Omega_{0} t$ is the expected evolution of the real part of the field,
and the wavevector has components ${\bf k}(t) = (-2 +
q\Omega_{0}t,1,1)(2\pi/L)$.  Note the highest resolution solution,
which has 32 grid points
per wavelength in each direction initially,
is indistinguishable from the semi-analytic solution.

It is instructive to compare the nonlinear saturation of the MRI in
calculations both with and without orbital advection.  We present
the evolution of two different three-dimensional calculations, one
with orbital advection and the other without.  Both start with no net
vertical flux, 
${\bf B} = (0,0,(2p_{0}/\beta)^{1/2}\sin{2\pi x/L_x})$,
where $\beta=400$, use a domain of size $L_{x}=L_{y}=8H$, and $L_{z}=H$
(where $H=C_{s}/\Omega_{0}=1$), and adopt an isothermal equation of
state with $C_{s}=10^{-3}$.  The resolution is $32/H$ in each dimension.
Figure 8 plots the time evolution of the magnetic energy
in both cases.  In the nonlinear regime,
the timestep in the calculation with orbital advection was on average 4.3
times larger than the calculation without.  The results show the solutions
are in excellent qualitative agreement.  Because the nonlinear evolution
of the MRI is chaotic (Winters et al. 2003), we do not expect exact
agreement, but only the same values for suitably time-averaged quantities.
The level of agreement between our two calculations is similar to that
reported by JGG.  These authors point out that because the truncation
error varies with $x$ in the shearing box without orbital advection,
there are unphysical patterns in the radial profile of time-averaged
quantities such as the stress and density.  Figure 9 plots these profiles
for this calculation.  Note that without orbital advection, we observe
a minimum in the density, and a maximum in the stress, near the center of
the calculation.  This is likely caused by a minimum in the truncation
error in the region of this point, since the velocity amplitude (including
the radial, azimuthal, and vertical components)
is a minimum near there.
However, with orbital advection the profiles of the density and stress
are clearly much smoother; they only
show small amplitude variations associated with the MRI-driven
turbulence.

These calculations do not use the corrections to the shearing box
boundary conditions (Gressel \& Ziegler 2007) that are required to
conserve mass and momentum exactly (energy is not conserved with an
isothermal equation of state).  Nonetheless, we find that at the end
of the calculation without orbital advection, mass is conserved to one
part in $10^{3}$, and with orbital advection it is conserved to one
part in $10^{5}$.  Since these simulations are only run for 16 orbits,
we have also checked mass conservation at various resolutions for zero
net vertical simulations in a domain of size $L_{x}=H$, $L_{y}=4H$, and
$L_{z}=H$, using resolutions of $32/H$, $64/H$, $128/H$, and $256/H$.
This study was conducted to confirm the lack of convergence of the
amplitude of the Maxwell stress in zero-net flux unstratified boxes
discussed by Fromang et al. (2007), and indeed we confirm this result.
A variety of other results from this study will be reported elsewhere.
Here we note that without orbital advection, mass conservation can be a
problem at the lowest resolutions.  At $32/H$, roughly 8\% of the initial
mass is lost by 100 orbits, with the mass loss rate essentially constant
with time.  The mass loss rate decreases with resolution, however even
at $256/H$, 1.5\% of the mass is lost by 100 orbits.  The use of orbital
advection greatly improves conservation.  After 100 orbits, we find only
0.03\% of the initial mass is lost by 100 orbits with orbital advection,
and this number is independent of resolution.

Finally, we have also used simulations of the MRI with an adiabatic
equation of state to investigate the implementation of the source terms
in the energy equation in the orbital advection step.  We have repeated
a simulations with net vertical flux ${\bf B} = (0,0,(2p_{0}/\beta)^{1/2})$,
where $\beta=400$, in a domain of size  $L_{x}=H$, $L_{y}=2\pi H$, and
$L_{z}=H$, using a resolution of $64 \times 128 \times 64$.  This 
is a repeat of run Z2 in Gardiner \& Stone (2005b).  The calculation
starts with $p_{0}=10^{-6}$, and uses $\gamma=5/3$.  We have run
calculations both with and without orbital advection, and compared the
resulting time evolution of the total energy with figure 3 in 
Gardiner \& Stone (2005b).  
Again, because of the chaotic nature of the MRI, we do not recover
exact agreement in the two calculations, however both show growth
of the total energy at very similar time-averaged rates, with all of the
increase after saturation occurring in the internal energy density.

\section{Application to the MRI}

The test results presented in the previous section demonstrate the orbital
advection algorithm is at least as accurate (although perhaps not any more
accurate) than integrations that do not use it.  The primary advantage
of orbital advection, however, is not that it is more accurate, but that
it removes the background shear flow from the time step stability limit,
and therefore enables much more efficient studies of accretion flows over
a wide range of radii.   Moreover, the new method based on CT introduced
in this paper is simpler than previous approaches, and preserves the
divergence free constraint to machine precision.

Figure 10 shows images of the
density and azimuthal velocity fluctuations in a shearing box simulation
of the MRI in a very wide domain of size $L_{x}=L_{y}=32H$, and $L_{z}=H$,
using a resolution of $32/H$ in each dimension, or $1024\times 1024\times
32$.  The initial conditions and parameters of the calculation are
identical to that presented in figure 8.  Interestingly the spiral density
waves excited by the MRI are immediately obvious, and have the same pitch
angle as in smaller domains, indicating the box size does not determine
these features (Heinemann \& Papaloizou 2008a; b).  Recently Johansen
et al. (2008) have reported long-lived density features in domains with
wide vertical extent, and a more careful analysis of this and other
runs reproduces this result.  The time averaged Maxwell stress in this
calculation is little different than the value in much smaller domains,
indicating the turbulent stress must be localized on scales $\lapprox H$
(Guan et al. 2009).

JGG found unphysical features in the time-averaged radial density profile
even with orbital advection, wherever the orbital shear displacement
was close to an integer number of zones.  In figure 11 we plot this
profile for the calculation shown in figure 10.  Using the time step
measured from this simulation, the radial locations where $\delta t
q\Omega_{0}/\delta y$ is an integer are plotted as vertical dashed lines.
There are no significant extrema at these locations, indicating the stress
and truncation error are smooth with radius.  This is another advantage
of the MHD orbital advection algorithm developed in this paper.

In summary, we have described the inclusion of source terms for
the shearing box approximation in the Athena MHD code, including
a Crank-Nicholson time differencing that preserves the amplitude of
epicyclic oscillations exactly.  We have also described an orbital
advection algorithm based on CT for evolving the induction equation
to preserve the divergence free constraint on the magnetic field.
We have shown this algorithm provides more accurate solutions at less
computational cost.  These algorithms have already been used to study
hydrodynamic turbulence in the shearing box (Shen et al. 2006).  A number
of new studies of the MRI in wide radial domains in both unstratified,
and vertically stratified disks are underway (Davis et al. 2010).

\acknowledgements
We thank Sebastien Fromang, Charles Gammie, John Hawley, Bryan Johnson,
E. Ostriker, and Jake Simon for helpful
discussions, and S. Fromang for providing the initial conditions
for the nonlinear density wave test, and B. Johnson for providing the reference
solution for the MHD shearing wave test.
Financial support was provided
by the DOE through DE-FG52-06NA26217.
Simulations were performed on
computational facilities at the Princeton Institute for Computational Science
and Engineering, and through resources provided by NSF grant AST-0722479.



\begin{figure}
\epsscale{0.8}
\plotone{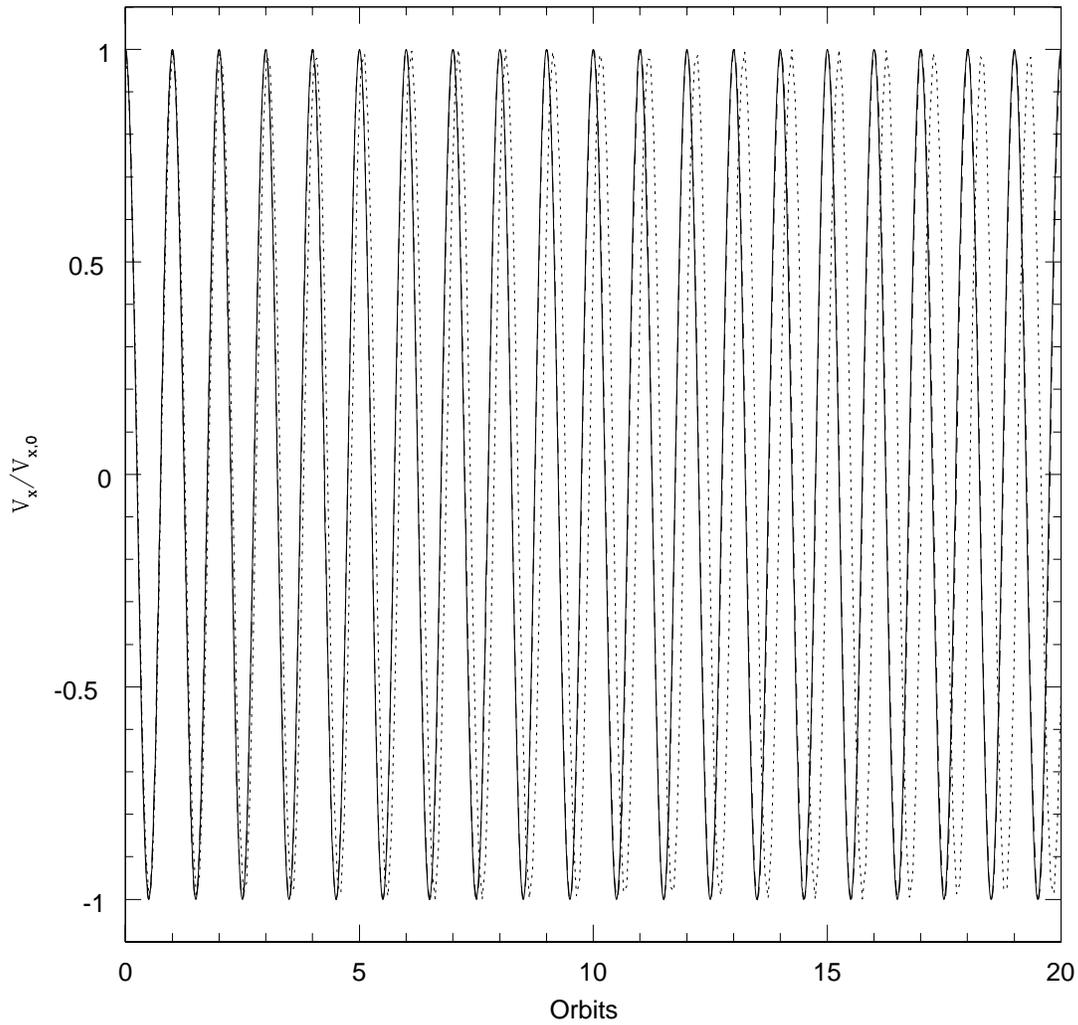}
\figcaption
{Time evolution of radial velocity amplitude, normalized by the initial
value, in epicyclic motion in a box of radial extent $L_{x}=1$ (solid
line),  $L_{x}=10$ (dashed line), and  $L_{x}=50$ (dotted line) computed
on a grid of $64^{2}$.
There is never amplitude error in the solutions, and only small
phase error in the largest box (in which the timestep is about 0.1 orbits).
}
\end{figure}

\begin{figure}
\epsscale{0.4}
\plotone{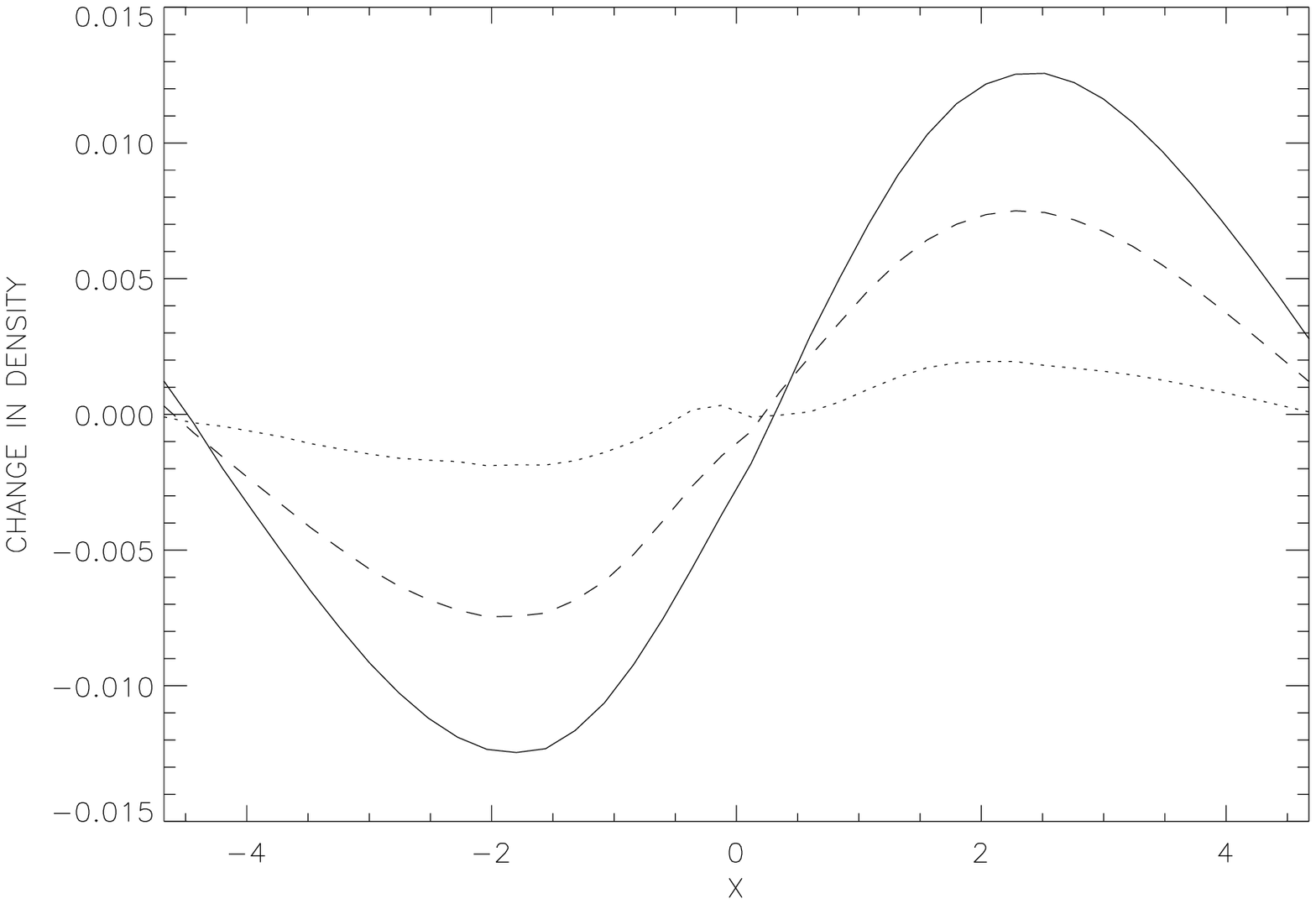}
\plotone{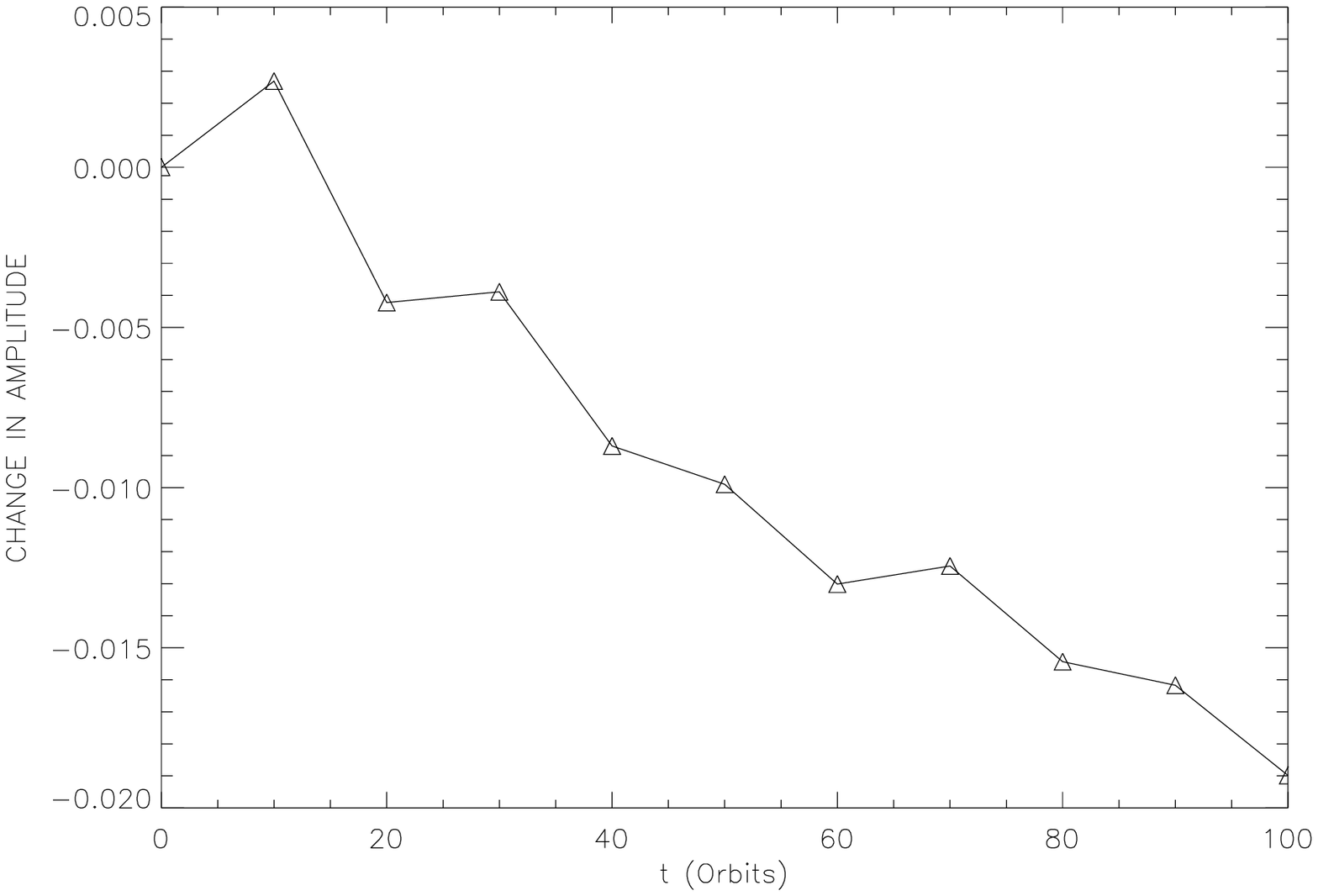}
\figcaption
{({\em Left}).  Fractional change in density in an axisymmetric nonlinear
density wave at $t=10T$ (dotted line), $50T$ (dashed line), and $100T$
(solid line), where $T=5254.31$ is the wave period, on a grid of 40 cells.
({\em Right}).  Fractional change in
the wave amplitude as a function of time.  Both quantities are defined
in the text.}
\end{figure}

\begin{figure}
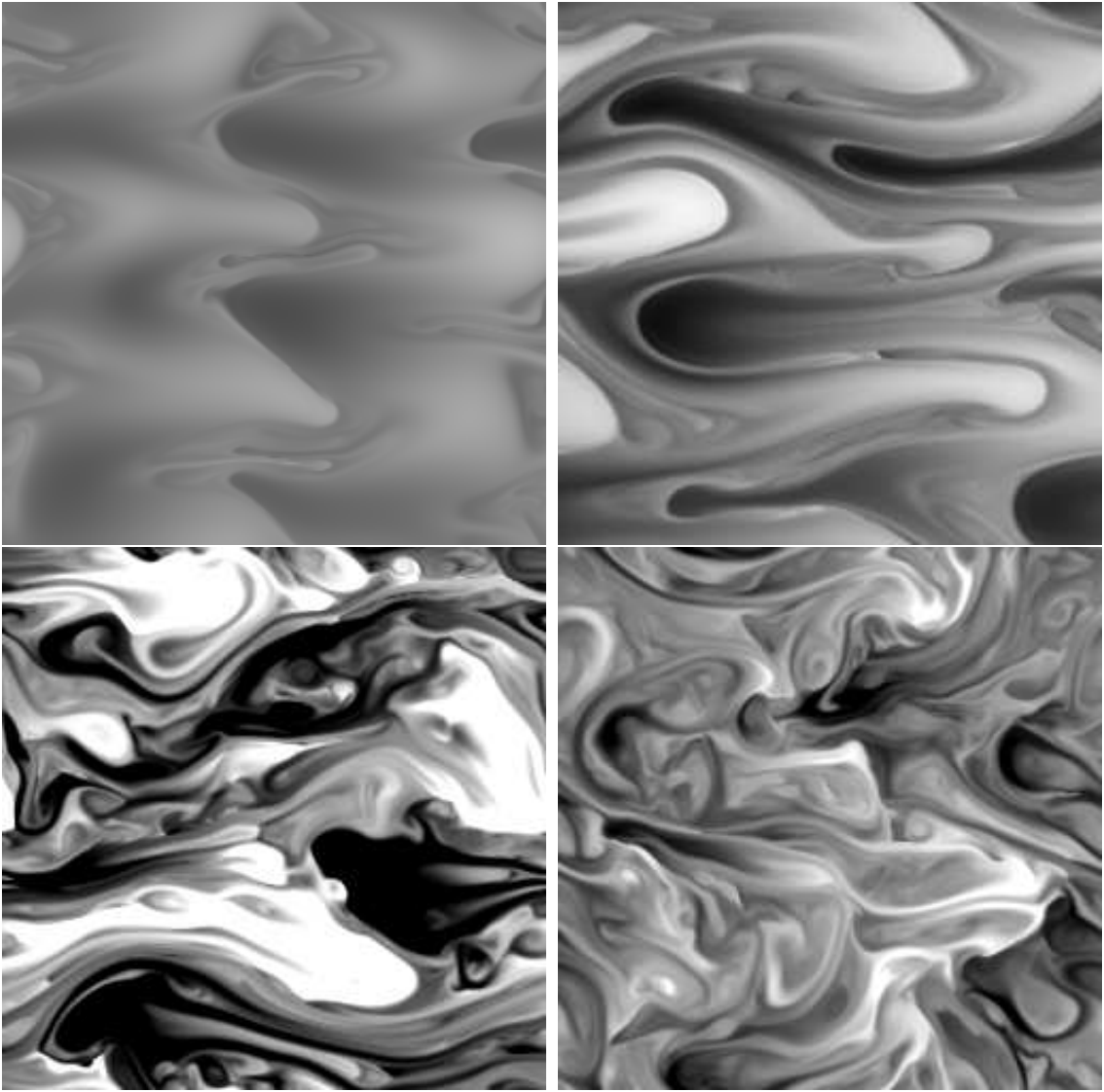

\epsscale{0.4}
\plotone{fig02a.ps3}
\plotone{fig02b.ps3}
\plotone{fig02c.ps3}
\plotone{fig02d.ps3}
\figcaption
{Images of the azimuthal velocity fluctuations $\delta v_{y} = v_{y} -
0.75\Omega_{0} x$ in the axisymmetric MRI computed with a grid of $256^{2}$.
From left-to-right and top-to-bottom the images are shown at times of 3.45, 
3.70, 4.0, and 6.0 orbits respectively.  The images are scaled over
$\pm 6 \times 10^{-4}$. }
\end{figure}

\begin{figure}
\epsscale{0.8}
\plotone{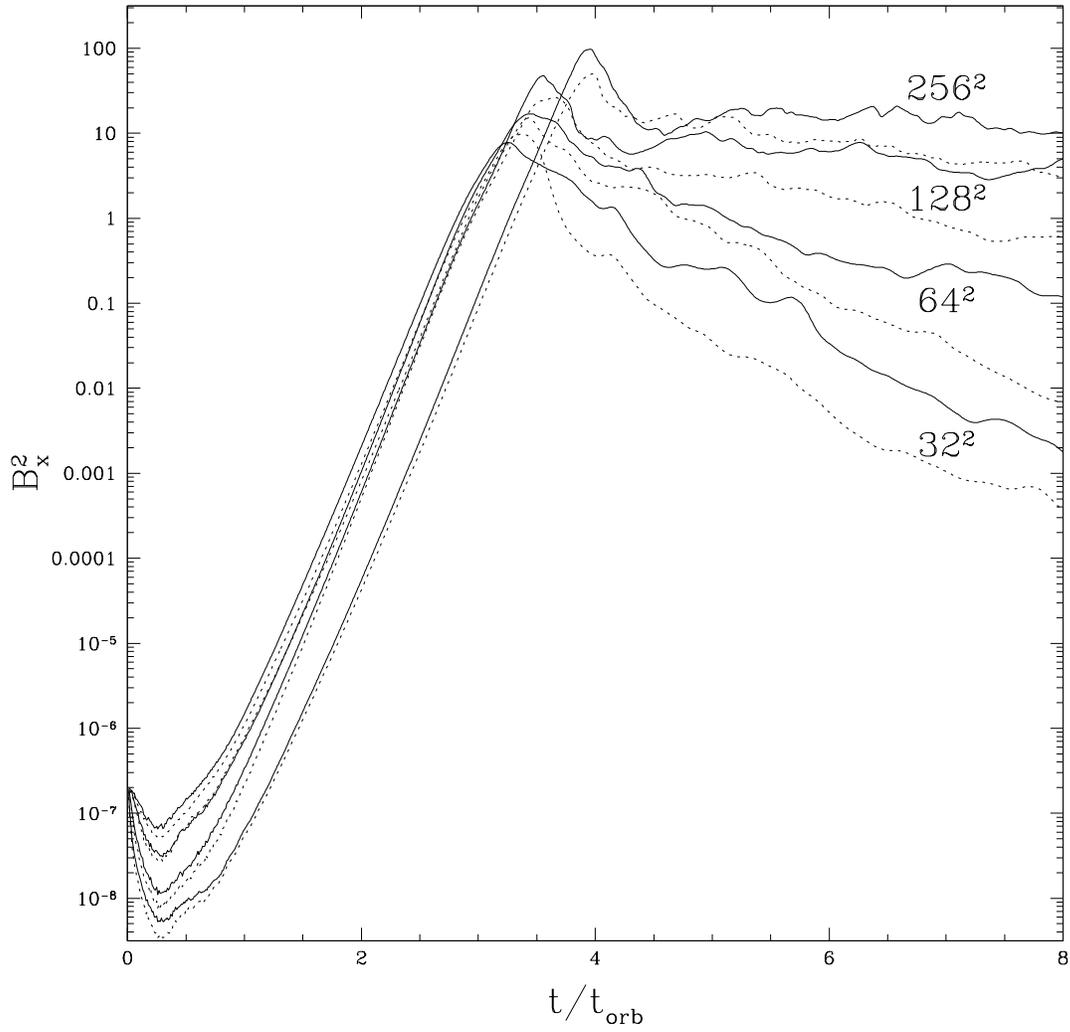}
\figcaption
{Time evolution of $0.5B_{x}^{2}$ in the axisymmetric MRI.  The solid lines
are for third-order spatial interpolation, the dashed lines are for
second-order.  The numerical resolution of each pair of lines is labeled. }
\end{figure}

\begin{figure}
\epsscale{0.4}
\plotone{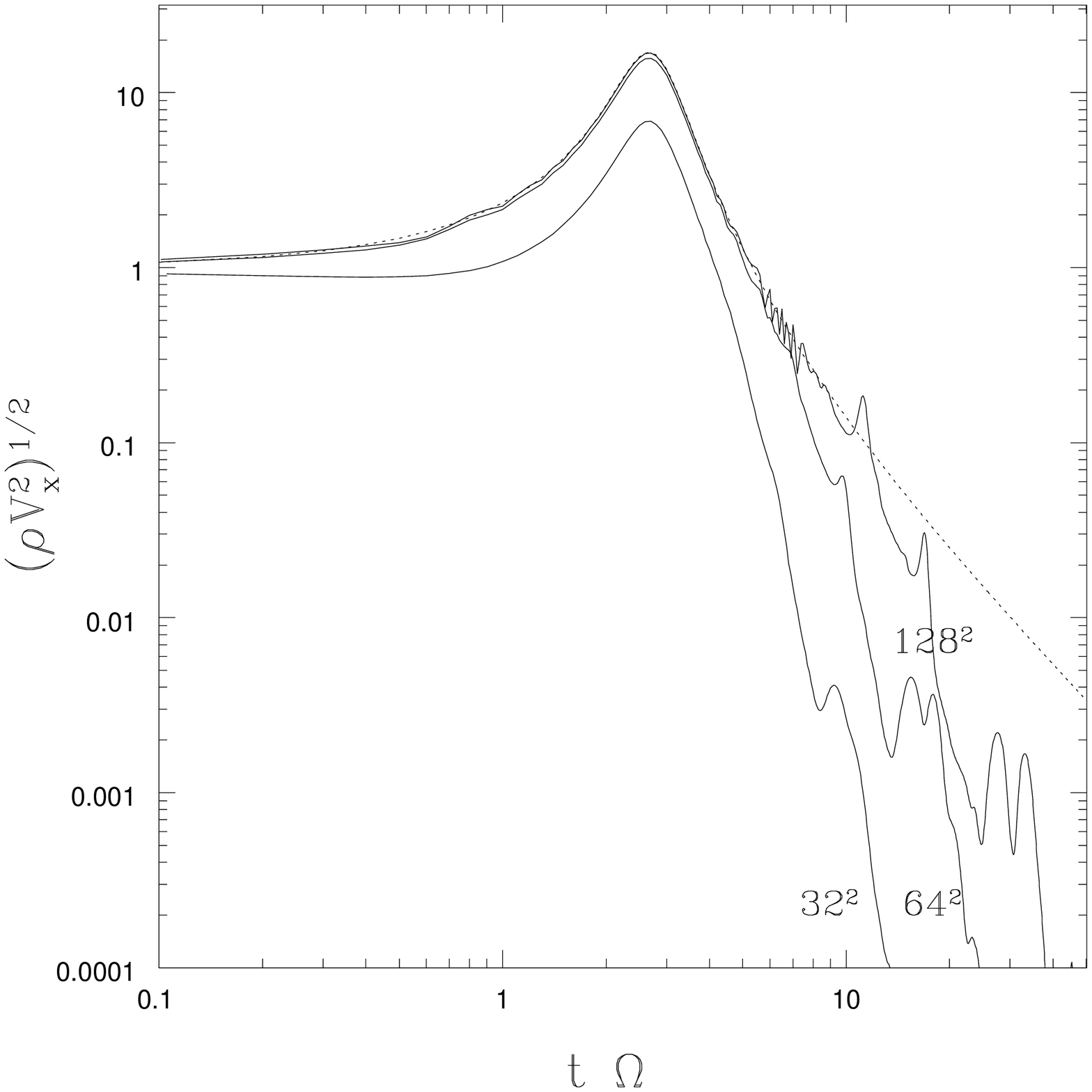}
\plotone{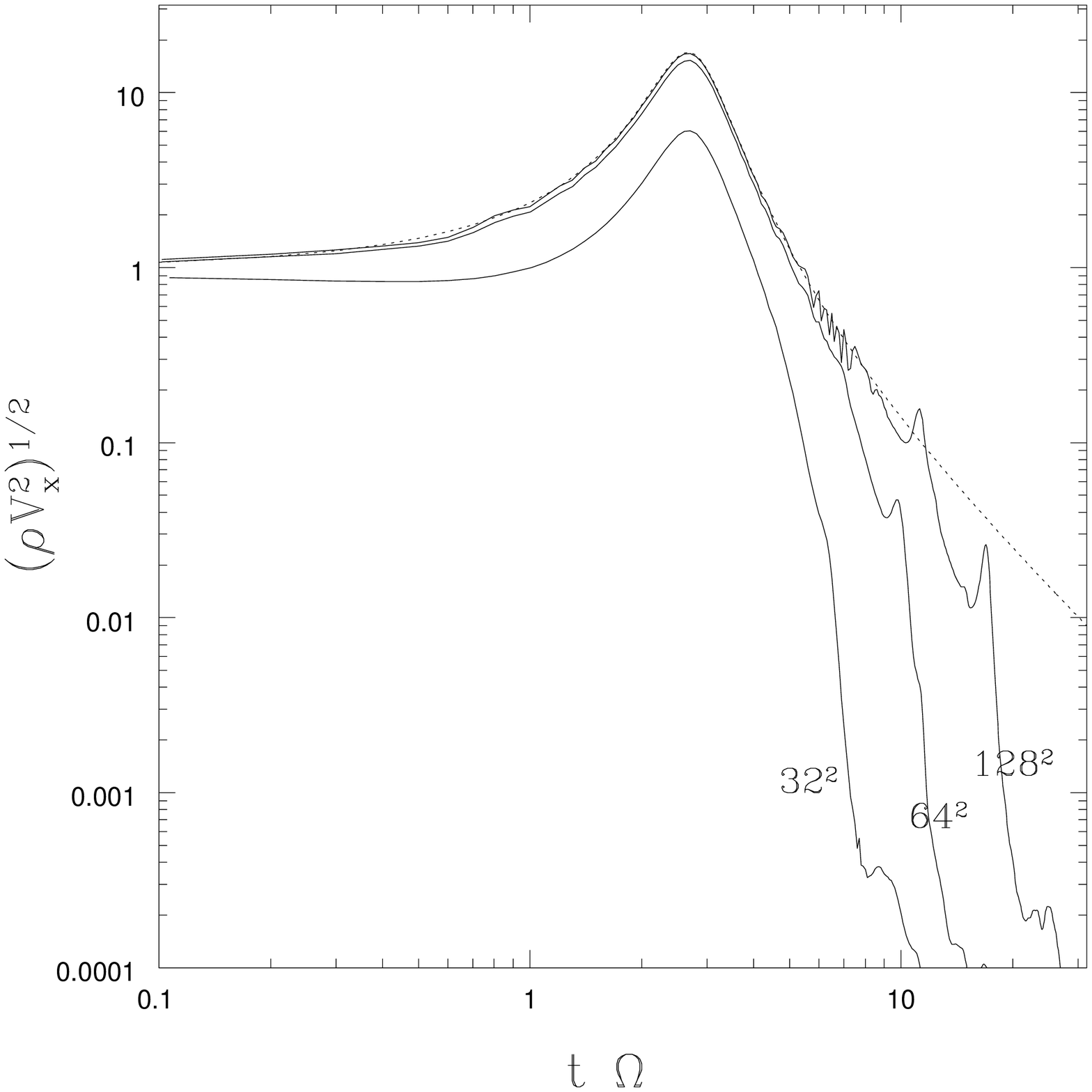}
\figcaption
{Time evolution of the radial velocity amplitude in an incompressible
shearing wave at numerical resolutions of $32^3$, $64^3$, and $128^2$,
corresponding to 4, 8, and 16 grid points per wavelength initially.
Solutions computed both without orbital advection {\em (left)},
and with orbital advection {\em (right)} are shown.
The dashed line in both cases is the analytic solution, and for each curve
the amplitude is
normalized by the initial value.  Particularly notable is the extremely
small level of aliasing.}
\end{figure}

\begin{figure}
\epsscale{0.2}
\plotone{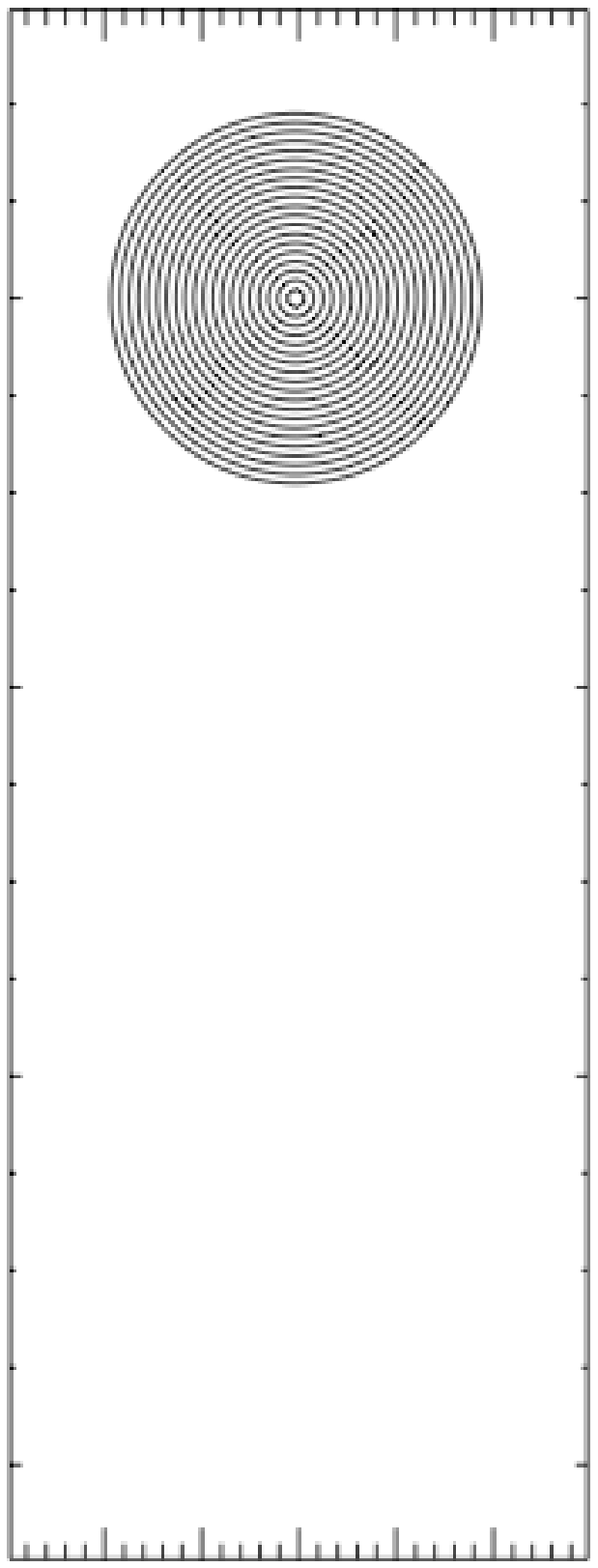}
\plotone{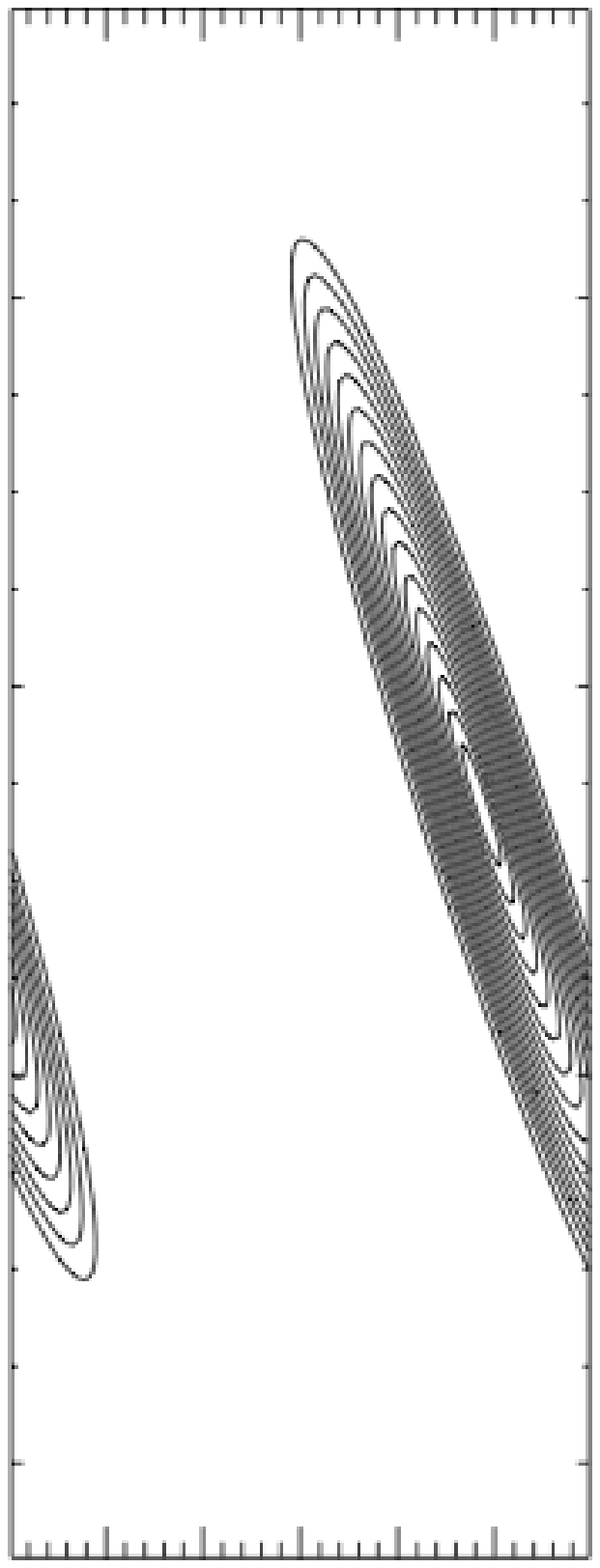}
\plotone{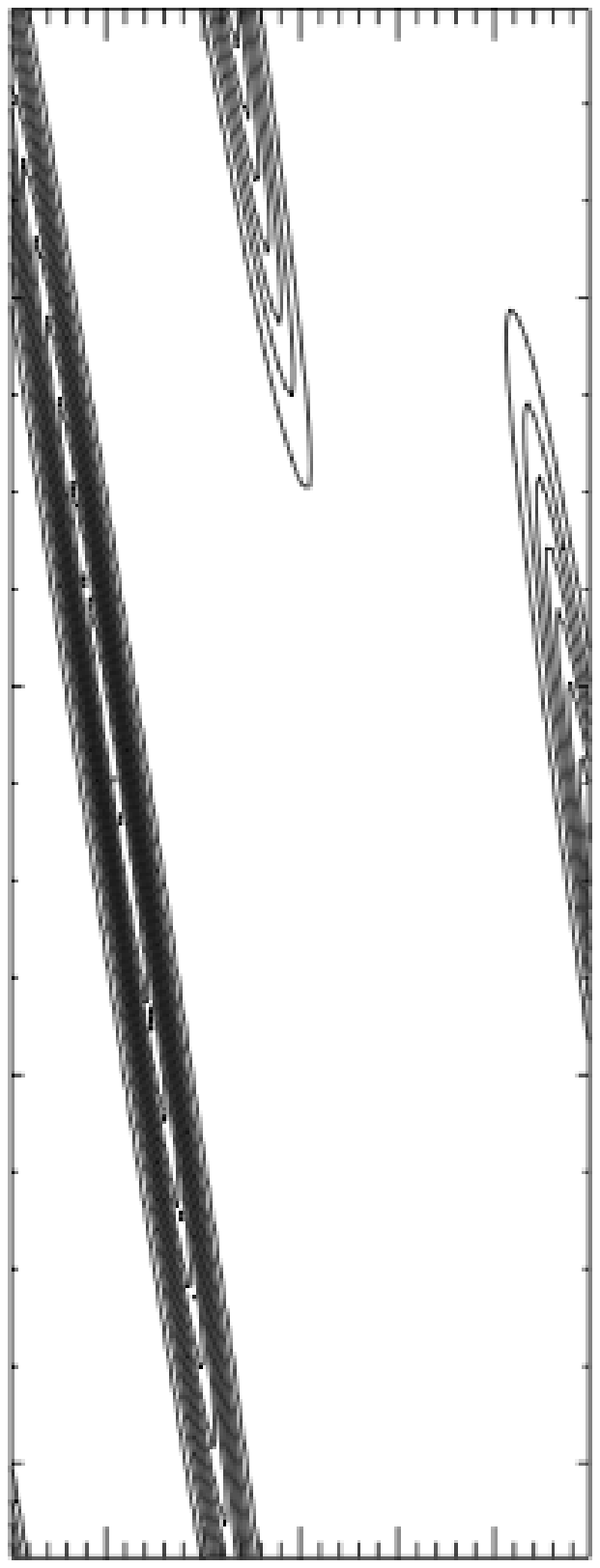}
\plotone{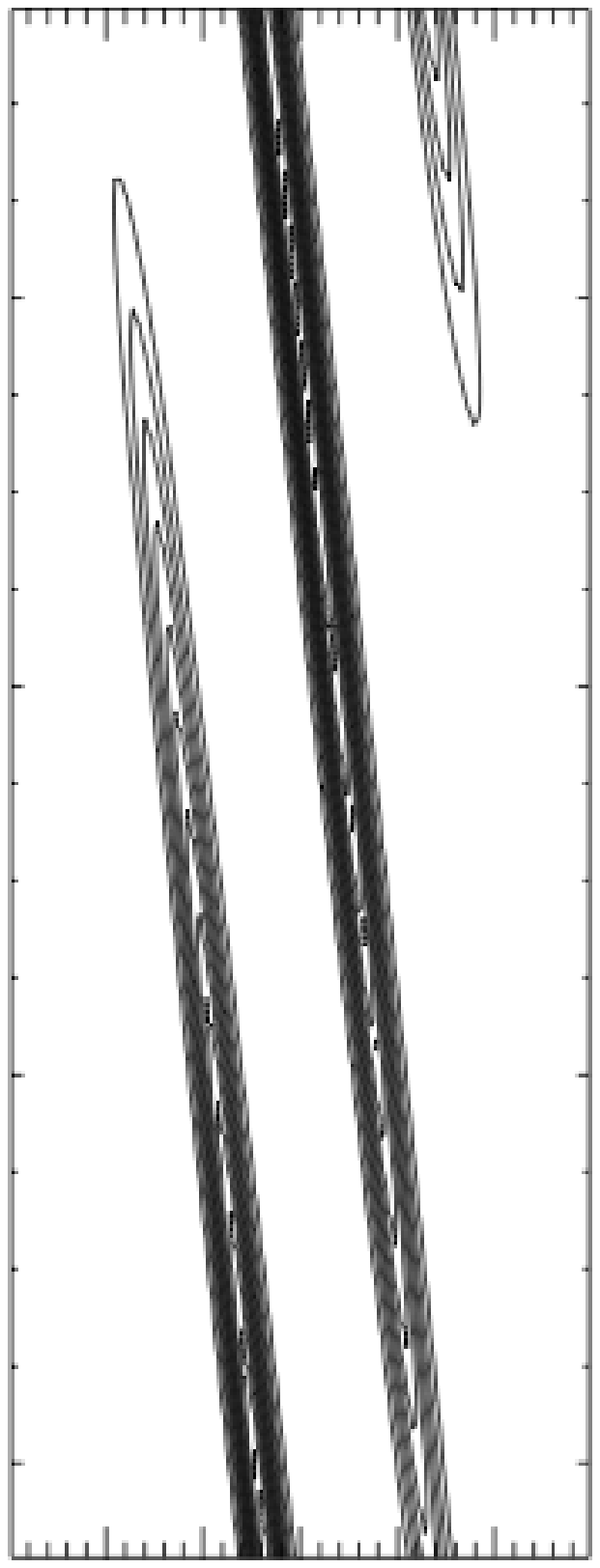}
\figcaption
{Field lines of an initially circular field loop advected by epicyclic motion
and sheared by the background flow, computed using orbital advection.
Twenty field
lines are shown at each times of (from left to right) $t=0$, 0.3, 0.7,
and 1 orbit.}
\end{figure}

\begin{figure}
\epsscale{0.8}
\plotone{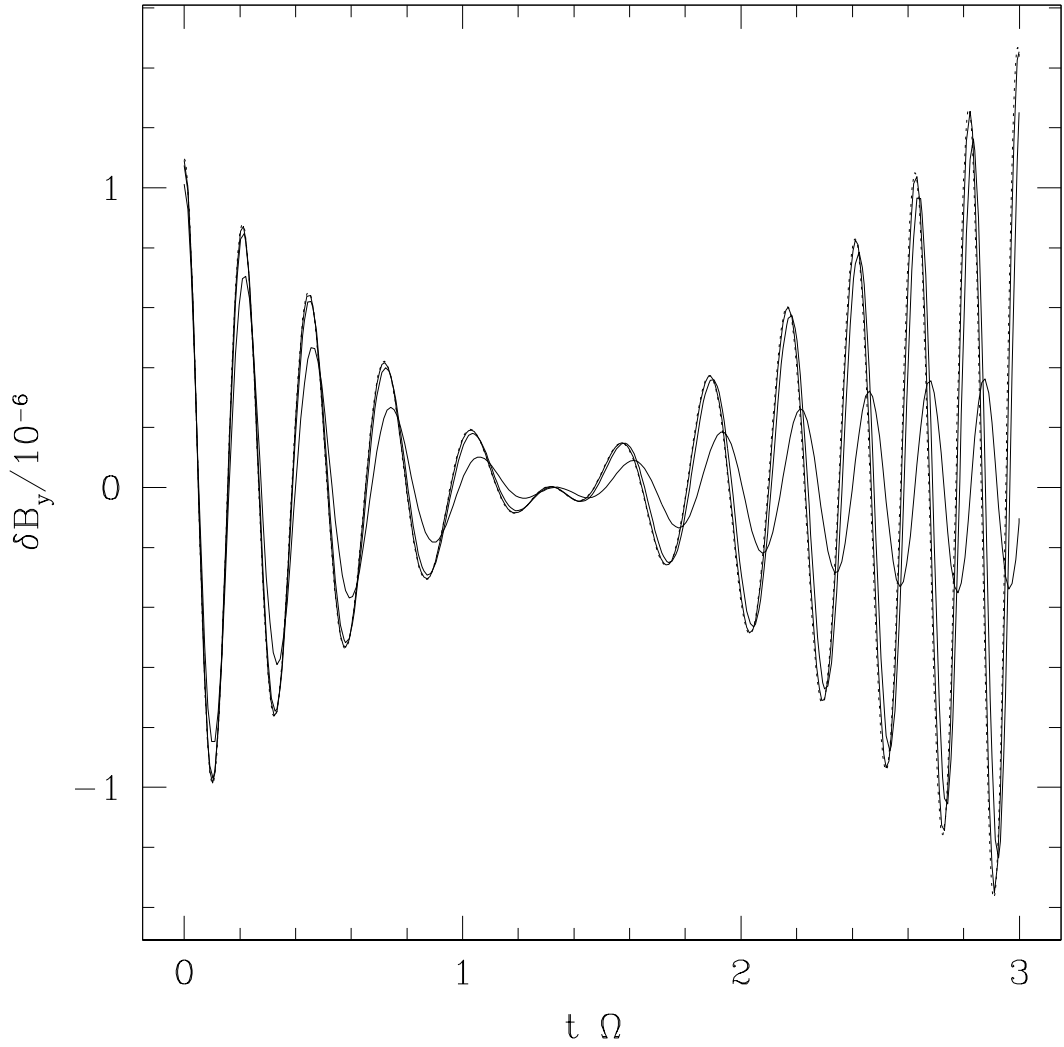}
\figcaption
{Time evolution of the real part of the wave amplitude in
the azimuthal component of the magnetic field for an MHD shearing wave test,
computed
using orbital advection.  The solid lines are at a resolution of 8, 16, and
32 grid points per $L_{z}$, while the dashed line is the semi-analytic
solution.}
\end{figure}

\begin{figure}
\epsscale{0.8}
\plotone{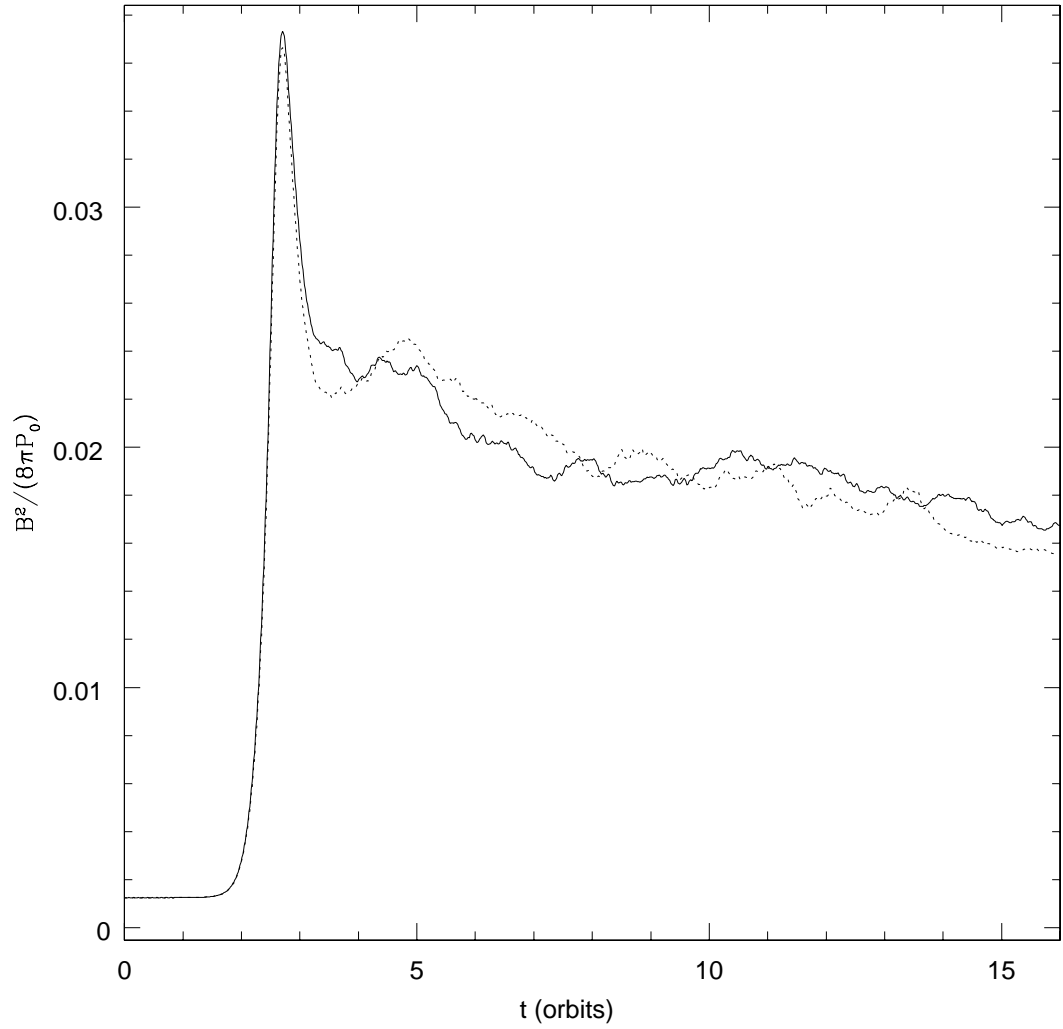}
\figcaption
{Time evolution of the magnetic energy in a zero-net flux MRI calculation
in a box with radial dimension $L_{x}=8H$.  The solid line is computed
with orbital advection, the dashed without.}
\end{figure}

\begin{figure}
\epsscale{0.4}
\plotone{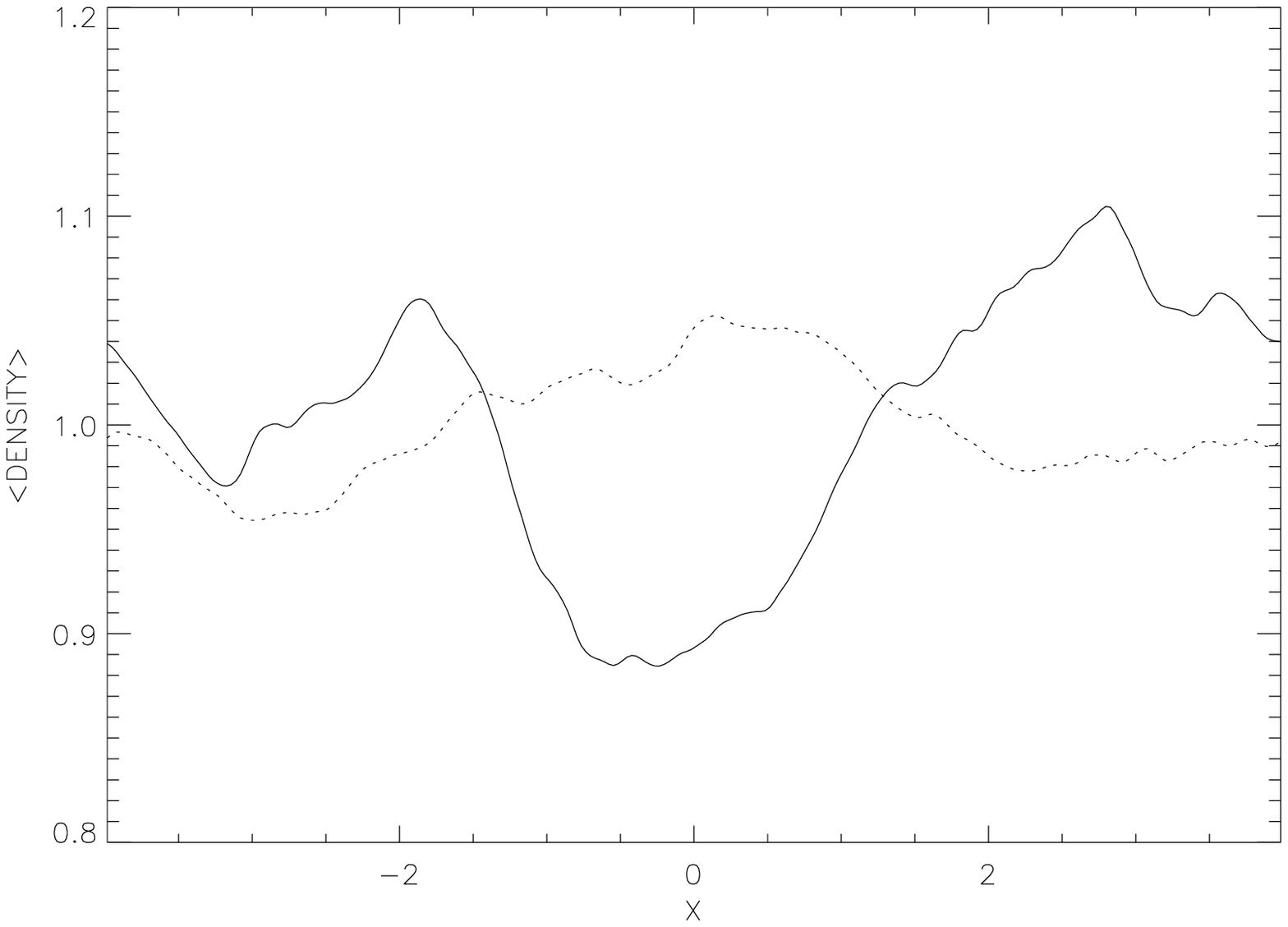}
\plotone{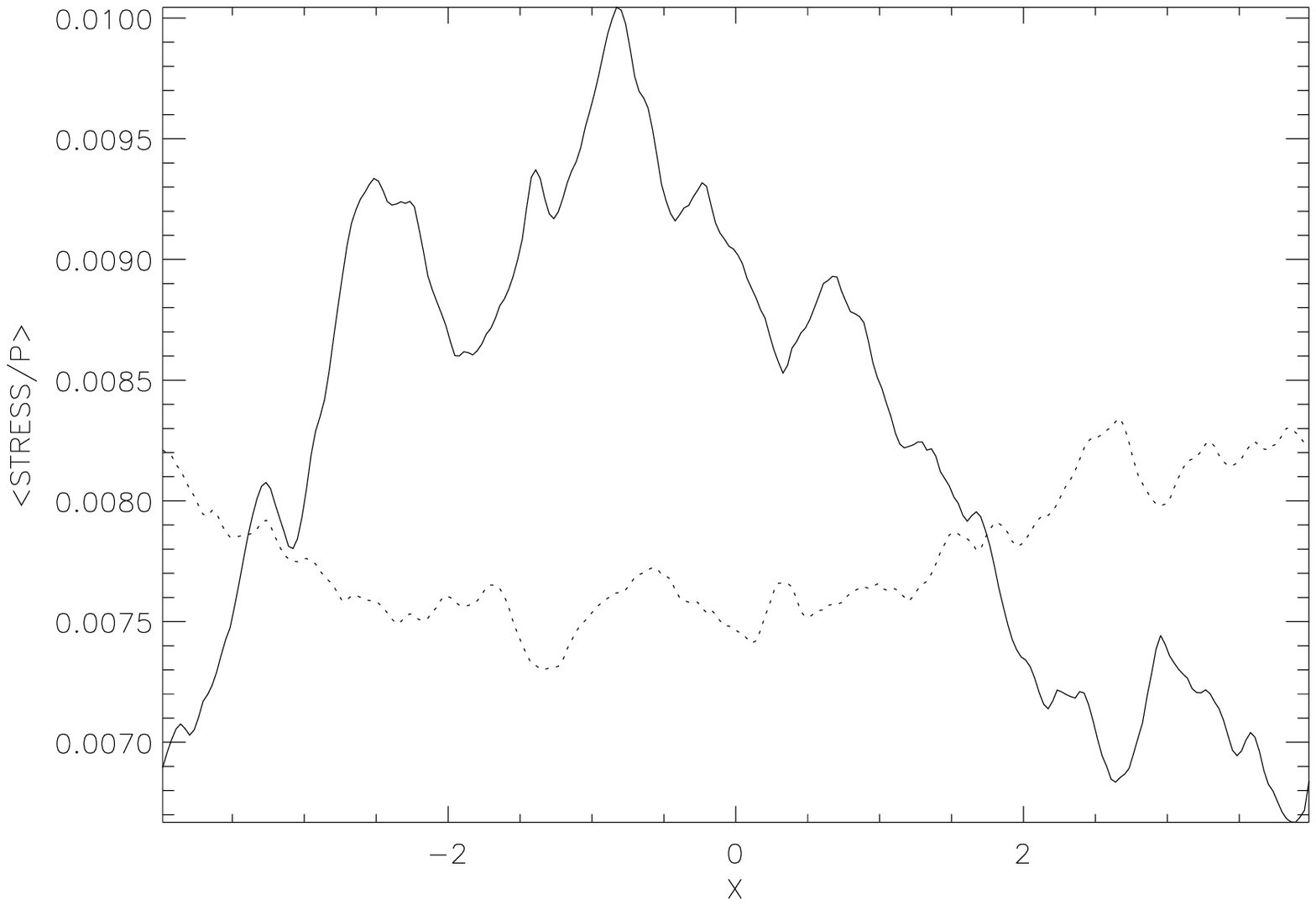}
\figcaption
{Radial profile of the average density (left) and stress (right) in
a zero-net flux MRI calculation computed with (dashed line) and without
(solid line) orbital advection.  The averages are taken over all $y$ and $z$,
and over orbits 7-16 for the solid line, and 70-100 for the dashed line.}
\end{figure}

\begin{figure}
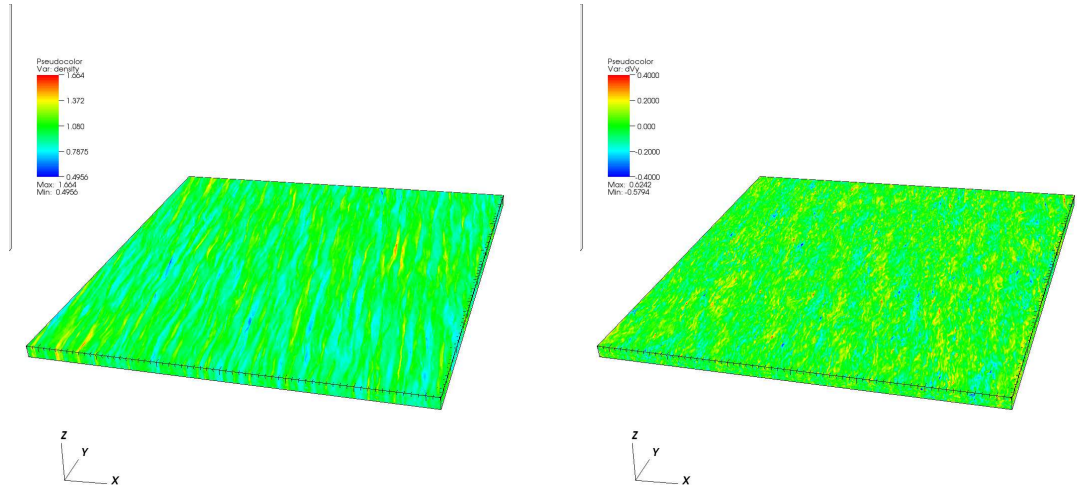

\epsscale{0.4}
\plotone{fig10a.ps3}
\plotone{fig10b.ps3}
\figcaption
{Images of the density (left) and azimuthal velocity fluctuations scaled to
the sound speed (right) from a zero-net flux MRI calculation in a box with
radial dimension $L_{x}=32H$ at $t=16$ orbits, computed with orbital advection.}
\end{figure}

\begin{figure}
\epsscale{0.8}
\plotone{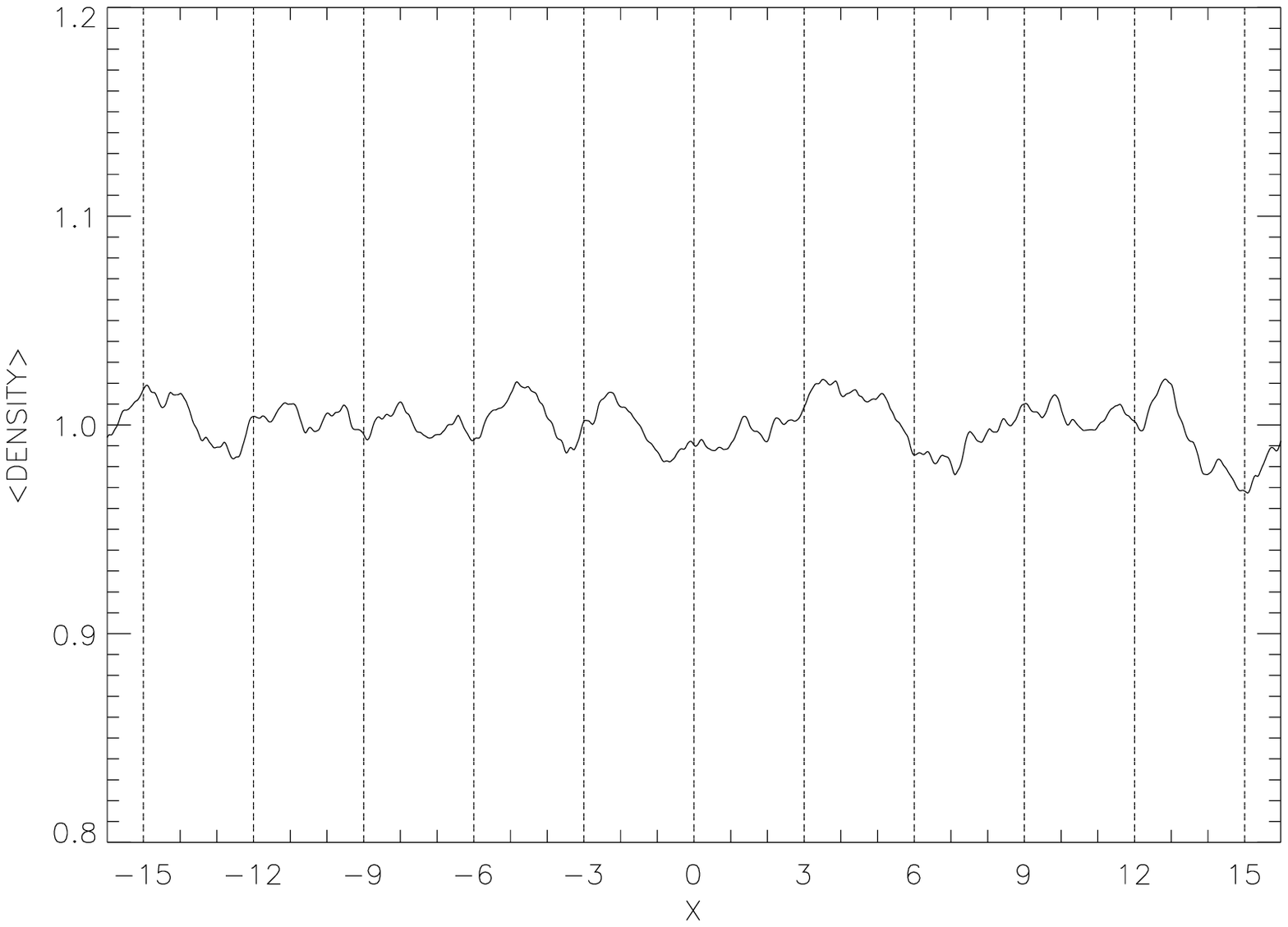}
\figcaption
{Radial profile of the average density in
a zero-net flux MRI calculation in a box with
radial dimension $L_{x}=32H$. 
The average is taken over all $y$ and $z$,
and over orbits 7-16.  For the typical timestep used in the calculation the
azimuthal displacement in the orbital advection algorithm is an integer number of zones at
the locations marked by the vertical dashed lines.  There is no evidence
of density minima at these locations.}
\end{figure}


\begin{references}

\reference{} Balbus, S.A., \& Hawley, J.F., 1991.  ApJ, 376, 214.

\reference{} Balbus, S.A., \& Hawley, J.F., 2003. LNP, 614, 329.

\reference{} Balbus, S.A. 2003.  ARA\&A, 41, 555.

\reference{} Balbus, S.A., \& Hawley, J.F., 2006.  ApJ 652, 1020.

\reference{} Bodo, G., Mignone, A., Cattaneo, F., Rossi, P., \& Ferrari, A.  
2008.  A\&A 487, 1.

\reference{} Davis, S.W., Stone, J.M., \& Pessah, M., 2010.  ApJ 713, 52.

\reference{} Evans, C.R., \& Hawley, J.F., 1988. Ap.J., 322, 659

\reference{} Fromang, S., \& Papaloizou,J., 2007.  A\& A, 468, 1.

\reference{} Fromang, S., Papaloizou, J., Lesur, G., \& Heinemann, T. 2007.  
A\&A 476, 1123

\reference{} Gardiner, T.A., \& Stone, J.M., 2005a. JCoPh, 205, 509

\reference{} Gardiner, T.A., \& Stone, J.M., 2005b. In {\em Magnetic Fields in
the Universe}, AIP Conf. Proc. 784, 475

\reference{} Gardiner, T.A., \& Stone, J.M., 2008. JCoPh, 227, 4123.

\reference{} Gressel, O., \& Ziegler, U. 2007.  CoPhC, 176, 652.

\reference{} Gressel, O., 2010.  MNRAS, in press.

\reference{} Guan, X., Gammie, C.F., Simon, J.B., \& Johnson, B.M. 2009.
ApJ 694, 1010.

\reference{} Hawley, J.F., \& Balbus, S.A., 1992.  ApJ 400, 595.

\reference{} Hawley, J.F., Gammie, C.F., \& Balbus, S.A., 1995.  ApJ 440, 742.

\reference{} Heinemann, T., \& Papaloizou, J.C.B., 2008a.  arXiv0812.2068.

\reference{} Heinemann, T., \& Papaloizou, J.C.B., 2008b.  arXiv0812.2471.

\reference{} Hill, G.W., 1878.  Amer. J. Math. 1, 5.

\reference{} Hirose, S., Krolik, J., \& Blaes, O., 2009.  ApJ 691, 16.

\reference{} Ilgner, M., \& Nelson, R.P. 2008.  A\&A, 483, 815.

\reference{} Johansen, A., Youdin, A., \& Klahr, H. 2009.  ApJ 697, 1269.

\reference{} Johnson, B.M., \& Gammie, C.F. 2005.  ApJ 626, 978.

\reference{} Johnson, B.M. 2007.  ApJ 660, 1375.

\reference{} Johnson, B.M., Guan, X., \& Gammie, C.F. 2008a.  ApJS 177, 373.
Addendum: ApJS 179, 553.

\reference{} Lesur, G., \& Longaretti, P.-Y., 2007.  A\&A, 378, 1471.

\reference{} Masset, F., 2000.  A\& AS, 141, 165.

\reference{} Piontek, R.A., Gressel, O., \& Ziegler, U. 2009.  A\& A 499, 633.

\reference{} Regev, O., \& Umurhan, O.M., 2008.  A\& A, 481, 21.

\reference{} Shen, Y., Stone, J.M., \& Gardiner, T.A., 2006.  ApJ 653, 513.

\reference{} Simon, J.B., \& Hawley, J.F., 2009.  ApJ 707, 833.

\reference{} Simon, J.B., Hawley, J.F., \& Beckwith, K., 2009.  ApJ 690, 974.

\reference{} Stone, J.M., Hawley, J.F., Gammie, C.F., \& Balbus, S.A. 1996.
ApJ 463, 656.

\reference{} Stone, J.M. \& Gardiner, T.A., 2005. In {\em Magnetic Fields in
the Universe}, AIP Conf. Proc. 784, 16

\reference{} Stone, J.M., Gardiner, T.A., Teuben, P., Hawley, J.F., \&
Simon, J.B., 2008. ApJS 178, 137 (SGTHS)

\reference{} Stone, J.M., \& Gardiner, T.A., 2009.  NewA 14, 139.

\reference{} Stone, J.M. 2009.  ASP Conf. Proc. 406, 277.

\reference{} Tilley, D.A., Balsara, D.S., Brittain, S.D., \& Rettig, T., 2010.
MNRAS 403, 211.

\reference{} Turner, N.J., \& Sano, T., 2008.  ApJ 679, L131.

\reference{} Winters, W.F., Balbus, S.A., \& Hawley, J.F. 2003.  MNRAS 340, 519.

\end{references}
\end{document}